\documentclass[fleqn,usenatbib]{mnras}

\usepackage{newtxtext,newtxmath}

\usepackage[T1]{fontenc}

\usepackage[usenames,dvipsnames]{xcolor}
\usepackage{placeins}
\usepackage{balance}

\DeclareRobustCommand{\VAN}[3]{#2}
\let\VANthebibliography\thebibliography
\def\thebibliography{\DeclareRobustCommand{\VAN}[3]{##3}\VANthebibliography}

\usepackage{graphicx}	
\usepackage{amsmath}	
\usepackage{orcidlink}

\hypersetup{
    colorlinks = true,
    citecolor = {MidnightBlue},
    linkcolor = {BrickRed},
    urlcolor = {BrickRed}
}

\usepackage{caption}
\usepackage{subcaption}
\usepackage{comment}

\usepackage{tikz}
\usetikzlibrary{shapes.geometric, arrows}

\definecolor{cinput}{HTML}{d9edf8}
\definecolor{cfilter}{HTML}{f5a59a}
\definecolor{cstage}{HTML}{B5CCCF}

\title[HERA fringe-rate filters for delay spectra]{A demonstration of the effect of fringe-rate filtering in the Hydrogen Epoch of Reionization Array delay power spectrum pipeline}

\author[H.Garsden et al.]{Hugh  Garsden\,\orcidlink{0009-0001-3949-9342}$^{1,2}$\thanks{Email: hugh.garsden@manchester.ac.uk},
	Philip  Bull\,\orcidlink{0000-0001-5668-3101}$^{2,3}$,
	Michael J. Wilensky\,\orcidlink{0000-0001-7716-9312}$^{2}$,
	Zuhra  Abdurashidova$^{4}$,
	Tyrone  Adams$^{5}$,
\newauthor
	James E. Aguirre\,\orcidlink{0000-0002-4810-666X}$^{6}$,
	Paul  Alexander$^{7}$,
	Zaki S. Ali$^{4}$,
	Rushelle  Baartman$^{5}$,
	Yanga  Balfour$^{5}$,
\newauthor
	Adam P. Beardsley\,\orcidlink{0000-0001-9428-8233}$^{8,9}$,
	Lindsay M. Berkhout\,\orcidlink{0000-0002-2293-9639}$^{8}$,
	Gianni  Bernardi\,\orcidlink{0000-0002-0916-7443}$^{10,11,5}$,
	Tashalee S. Billings$^{6}$,
\newauthor
	Judd D. Bowman\,\orcidlink{0000-0002-8475-2036}$^{8}$,
	Richard F. Bradley$^{12}$,
	Jacob  Burba\,\orcidlink{0000-0002-8465-9341}$^{2}$,
	Steven  Carey$^{7}$,
	Chris L. Carilli\,\orcidlink{0000-0001-6647-3861}$^{13}$,
\newauthor
	Kai-Feng  Chen\,\orcidlink{0000-0002-3839-0230}$^{14,15}$,
	Carina  Cheng$^{4}$,
	Samir  Choudhuri$^{1,16}$,
	David R. DeBoer\,\orcidlink{0000-0003-3197-2294}$^{17}$,
	Eloy  de~Lera~Acedo$^{7}$,
\newauthor
	Matt  Dexter$^{17}$,
	Joshua S. Dillon\,\orcidlink{0000-0003-3336-9958}$^{4}$,
	Scott  Dynes$^{14}$,
	Nico  Eksteen$^{5}$,
	John  Ely$^{7}$,
\newauthor
	Aaron  Ewall-Wice\,\orcidlink{0000-0002-0086-7363}$^{4,18}$,
	Nicolas  Fagnoni$^{7}$,
	Randall  Fritz$^{5}$,
	Steven R. Furlanetto\,\orcidlink{0000-0002-0658-1243}$^{19}$,
\newauthor
	Kingsley  Gale-Sides$^{7}$,
	Bharat Kumar Gehlot$^{8}$,
	Abhik  Ghosh$^{3}$,
	Brian  Glendenning$^{20}$,
	Adelie  Gorce$^{21}$,
\newauthor
	Deepthi  Gorthi\,\orcidlink{0000-0002-0829-167X}$^{4}$,
	Bradley  Greig\,\orcidlink{0000-0002-4085-2094}$^{22}$,
	Jasper  Grobbelaar$^{5}$,
	Ziyaad  Halday$^{5}$,
	Bryna J. Hazelton\,\orcidlink{0000-0001-7532-645X}$^{23,24}$,
\newauthor
	Jacqueline N. Hewitt\,\orcidlink{0000-0002-4117-570X}$^{14,15}$,
	Jack  Hickish$^{17}$,
	Tian  Huang$^{7}$,
	Daniel C. Jacobs\,\orcidlink{0000-0002-0917-2269}$^{8}$,
	Alec  Josaitis$^{7}$,
\newauthor
	Austin  Julius$^{5}$,
	MacCalvin  Kariseb$^{5}$,
	Nicholas S. Kern\,\orcidlink{0000-0002-8211-1892}$^{14,15,\dagger}$,
	Joshua  Kerrigan\,\orcidlink{0000-0002-1876-272X}$^{25}$,
\newauthor
	Honggeun  Kim\,\orcidlink{0000-0001-5421-8927}$^{14,15}$,
	Piyanat  Kittiwisit\,\orcidlink{0000-0003-0953-313X}$^{3}$,
	Saul A. Kohn\,\orcidlink{0000-0001-6744-5328}$^{6}$,
	Matthew  Kolopanis\,\orcidlink{0000-0002-2950-2974}$^{8}$,
	Adam  Lanman$^{25}$,
\newauthor
	Paul  La~Plante\,\orcidlink{0000-0002-4693-0102}$^{4,6}$,
	Adrian  Liu\,\orcidlink{0000-0001-6876-0928}$^{4,21}$,
	Anita  Loots$^{5}$,
	Yin-Zhe  Ma\,\orcidlink{0000-0001-8108-0986}$^{26}$,
	David H.~E. MacMahon$^{17}$,
\newauthor
	Lourence  Malan$^{5}$,
	Cresshim  Malgas$^{5}$,
	Keith  Malgas$^{5}$,
	Bradley  Marero$^{5}$,
	Zachary E. Martinot$^{6}$,
\newauthor
	Andrei  Mesinger\,\orcidlink{0000-0003-3374-1772}$^{27}$,
	Mathakane  Molewa$^{5}$,
	Miguel F. Morales\,\orcidlink{0000-0001-7694-4030}$^{23}$,
	Tshegofalang  Mosiane$^{5}$,
\newauthor
	Steven G. Murray\,\orcidlink{0000-0003-3059-3823}$^{27,8}$,
	Abraham R. Neben\,\orcidlink{0000-0001-7776-7240}$^{14,15}$,
	Bojan  Nikolic$^{7}$,
	Chuneeta Devi Nunhokee\,\orcidlink{0000-0002-5445-6586}$^{28,29}$,
\newauthor
	Hans  Nuwegeld$^{5}$,
	Aaron R. Parsons\,\orcidlink{0000-0002-5400-8097}$^{4}$,
	Robert  Pascua\,\orcidlink{0000-0003-0073-5528}$^{4,21}$,
	Nipanjana  Patra\,\orcidlink{0000-0002-9457-1941}$^{4}$,
	Samantha  Pieterse$^{5}$,
\newauthor
	Yuxiang  Qin$^{30,27}$,
	Eleanor  Rath$^{14,15}$,
	Nima  Razavi-Ghods$^{7}$,
	Daniel  Riley$^{14}$,
	James  Robnett$^{13}$,
\newauthor
	Kathryn  Rosie$^{5}$,
	Mario G. Santos$^{5,3}$,
	Peter  Sims\,\orcidlink{0000-0002-2871-0413}$^{21}$,
	Saurabh  Singh\,\orcidlink{0000-0001-7755-902X}$^{21,31}$,
	Dara  Storer$\,\orcidlink{0000-0003-4092-0103}^{23}$,
\newauthor
	Hilton  Swarts$^{5}$,
	Jianrong  Tan$^{6}$,
	Nithyanandan  Thyagarajan\,\orcidlink{0000-0003-1602-7868}$^{32,13}$,
	Pieter  van~Wyngaarden$^{5}$,
\newauthor
	Peter K.~G. Williams\,\orcidlink{0000-0003-3734-3587}$^{33,34}$,
	Zhilei  Xu\,\orcidlink{0000-0001-5112-2567}$^{14}$,
	Haoxuan  Zheng$^{15}$
\\
Affiliations are listed in Appendix~\ref{sec:affiliations}.
\vspace{-0.5cm}
}

\date{Accepted XXX. Received YYY; in original form ZZZ}

\pubyear{2024}

\begin{document}
\pagerange{\pageref{firstpage}--\pageref{lastpage}}
\maketitle

\begin{abstract}
Radio interferometers targeting the 21cm brightness temperature fluctuations at high redshift are subject to systematic effects that operate over a range of different timescales. These can be isolated by designing appropriate Fourier filters that operate in fringe-rate (FR) space, the Fourier pair of local sidereal time (LST). Applications of FR filtering include separating effects that are correlated with the rotating sky vs. those relative to the ground, down-weighting emission in the primary beam sidelobes, and suppressing noise. FR filtering causes the noise contributions to the visibility data to become correlated in time however, making interpretation of subsequent averaging and error estimation steps more subtle. In this paper, we describe fringe rate filters that are implemented using discrete prolate spheroidal sequences, and  designed for two different purposes -- beam sidelobe/horizon suppression (the `mainlobe' filter), and ground-locked systematics removal (the `notch' filter). We apply these to simulated data, and study how their properties affect visibilities and power spectra generated from the simulations. Included is an introduction to fringe-rate filtering and a demonstration of fringe-rate filters applied to simple situations to aid understanding.
\end{abstract}

\label{firstpage}


\clearpage

\section{Introduction}

Fluctuations in the brightness temperature of redshifted 21cm line emission from neutral hydrogen provide a unique probe of cosmic structure in the early Universe \citep{1990MNRAS.247..510S, 1993MNRAS.265..101S, furlanetto06, morales10, pritchard12, mellema13, 2014ApJ...782...66P}, from the cosmic `Dark Ages' following last-scattering ($z \approx 30 - 1000$), through Cosmic Dawn ($z \lesssim 30$) and the Epoch of Reionisation ($z \approx 6 - 15$). Other probes, such as the optical, IR, and UV emission from early stars and galaxies \citep{2016arXiv160607039D, 2022ARA&A..60..121R}, and molecular line emission \citep{2017arXiv170909066K}, suffer from obscuration by dust \citep{2021Natur.597..489F}, interlopers/confusion \citep{2016ApJ...825..143L, Cheng:2016yvu, 2020ApJ...894..152G, 2022MNRAS.512.4262C}, and more restrictive limits to their detectable redshift range. Developing practical approaches to measuring the 21cm fluctuations is therefore of great utility to the cosmology and galaxy formation communities. Despite over a decade of concerted effort using multiple generations of radio telescopes, a robust detection of this signal at high redshift ($z \gtrsim 6$) has remained elusive  \citep{2019arXiv190306218M, 2022arXiv220307864L}. The principal reason is the large dynamic range between the 21cm signal itself, which is rather faint (typically of order mK), and the radio emission from other (foreground) sources, measuring from tens of K at $\sim$GHz frequencies to thousands of K at $\lesssim 100$~MHz \citep{2005ApJ...625..575S}.

In principle, the 21cm fluctuations and the foreground sources have highly distinctive spectral and angular signatures, making it possible to separate them by applying appropriate cuts or filters in these domains \citep{2005ApJ...625..575S}. In practice, however, the imperfect calibration of any given radio telescope gives rise to spurious couplings and modulation of the signals, blurring the distinction between them \citep{2012ApJ...752..137M}. Since the foregrounds are so much brighter, even very small calibration errors can leak spurious foreground emission into otherwise 21cm-dominated regions of the data, so that recovery of the 21cm signal is severely hampered.

A wide variety of signal processing methods have been used to try and make headway with this problem. A very common approach is to transform the frequency-dependent data into a spectral basis, such as a Fourier basis or principal components of the data's frequency-frequency covariance matrix \citep[e.g.][]{2009MNRAS.398..401L, 2012ApJ...756..165P, 2014MNRAS.441.3271W, 2015MNRAS.447..400A}. This exploits the differing spectral behaviours (and different brightness/signal-to-noise) of the 21cm signal and foregrounds to provide some degree of separation between them, albeit imperfectly due to coupling/modulation effects mentioned above. For individual interferometer baselines, this can be thought of as applying a filter in the delay ($\tau$) domain, where delay is the Fourier pair of frequency, $\nu$, with Fourier basis functions $e^{2\pi i \nu\tau}$. Filters to remove foreground contamination are typically most aggressive at low delay, $\tau \sim 0$, but may also affect higher delays. This leads to some partial suppression of the 21cm signal too (otherwise known as `signal loss').

The `foreground avoidance' approach uses filters that separate out the most contaminated regions of the spectrum, including the 21cm signal in those regions, permitting only the nominally less-contaminated modes to be used in analysis. Conversely, the `foreground removal' approach attempts to model and remove the contamination while leaving the 21cm signal as intact as possible. Both approaches have serious drawbacks; foreground avoidance is lossy, and can still allow some degree of contamination or statistical biases if couplings between foreground-dominated and 21cm signal-dominated modes are sufficiently large \citep{2016MNRAS.461.3135B, 2016MNRAS.456...66J, 2017MNRAS.470.1849E}, whereas foreground removal can introduce additional artifacts into the data unless the subtracted model is very accurate \citep{2016ApJ...819....8P, 2019A&A...631A..12O}.

The spectral dimension of the data is not the only one that can usefully differentiate between the different signal components however. Baseline-dependent filtering of visibilities (or angular filtering of reconstructed images) can help separate contamination from our Galaxy (mostly at large angular scales) from extragalactic emission (on all scales) \citep{2023MNRAS.522.1009C}. Particular weighting and filtering schemes can also be used at the calibration step to suppress spurious features caused by antennas or baselines that are more susceptible to them.

In this paper, we focus on the design and effects of filters in {\it fringe-rate} ($f$) space \citep{2003ASPC..306..109R, Parsons:2009ju, 2016ApJ...820...51P}. This is the Fourier pair of local sidereal time (LST, denoted by $t$ in this paper), which for a drift-scan telescope is a proxy for Right Ascension (terrestrial drift-scan telescopes observe a stripe of the sky at fixed Declination, with increasing RA as the Earth rotates). The Fourier transform of time sequences of visibilities was first used in \citet{1973MNRAS.165...25P} to distinguish between fringe-rates as part of a technique for phase correcting interferometers. In 2003,  filtering of fringe-rates was used by \citet{2003MNRAS.341.1057W} to remove the Sun and Moon from observations of Jupiter using the Very Small Array \citep{2005IAUS..201..512R}.  Fringe-rate filtering  was applied to observations from the Precision Array for Probing the Epoch of Reionization (PAPER), used for the generation of 21cm power spectrum and upper limits on the EoR  \citep{2014ApJ...788..106P, 2015ApJ...809...61A}. It is intended to be a used on observations from  the Hydrogen Epoch of Reionization Array \citep[HERA;][]{2017PASP..129d5001D} as we will discuss. Other telescopes are adopting the technique, for example the VLA has used fringe-rate filters for RFI excision \citep{2019RaSc...54.1002H}.

The rate at which sky sources pass through the fringes of an interferometer baseline depends on the baseline length, and orientation of the baseline relative to the source path over time.
Fringe-rate filters are typically applied to each baseline separately. Sources rotating with the sky pass through the fringe pattern of a baseline more rapidly for longer baseline lengths, which have fringes closer together; the  visibilities generated by a source will oscillate (in phase) faster in time for those baselines, and will appear at high fringe rates. Sources further from the celestial poles have faster apparent motion on the sky, and so will have higher fringe rates than sources closer to the poles. The fringe rate can also be positive or negative depending on which side of the pole a source is at (with respect to the centre of the primary beam), and tends to decrease as sources get closer to the local horizon due to projection effects. Importantly, this means that fringe rate can be used as a crude proxy for source position, making it possible to filter or down-weight emission in the sidelobes of the primary beam, which contaminate Fourier modes needed for detecting the 21cm signal \citep{2015ApJ...807L..28T}. Removal of sky locations from the primary beam is in effect ``beam sculpting'', demonstrated in \citet{2016ApJ...820...51P}. The ``mainlobe'' filter to be described below is an example of this. 

Another application of fringe-rate filtering is to separate sources of emission that are locked to the sky's rotation from those that are not \citep[e.g.][]{2019ApJ...884..105K, Kern+20}. Stationary terrestrial sources will not fringe for example, so should be found at around $f \approx 0$, whereas aircraft, satellites, and other transient phenomena might be expected to give rise to signals that pass through the fringe pattern much faster than sky rotation would allow. The ``notch'' filter, described below, removes fringe-rates around $f \approx 0$, and the mainlobe filter can be used to deal with non-sky high fringe-rates. These filters will also help to excise unwanted systematics that produce spurious fringe-rates,  particularly if the systematic  has a differentiated spectral behaviour that can also be isolated in delay space. Finally, the thermal noise contribution to the data changes randomly from one time sample to the next. For white noise, this gives rise to a signal that has similar power at all fringe rates. By filtering out fringe-rates outside the range corresponding to sky-locked signals like the 21cm emission, some of this noise can be suppressed.

In forthcoming analyses of data from HERA, fringe-rate filtering will be adopted as a standard part of the observation analysis pipeline. This was driven by the presence of mutual coupling systematics that are well-localised in delay-fringe-rate space, and demonstrations showing that fringe-rate filters are useful in mitigating these \citep{Kern+20}. This can make calibration easier; for example, \citet{2024arXiv240720923C} demonstrated  its use in suppressing calibration errors arising from poorly modeled diffuse emission, showing roughly an order of magnitude improvement in calibration fidelity. More general beam-sculpting, i.e. the mainlobe filter, along with  the notch filter,   present interesting options for enhancing the use of fringe-rate filters in HERA data processing. 

With the benefits of fringe-rate filtering also come a number of drawbacks. Applying any filter can lead to some degree of signal loss (of the 21cm signal),  and introduce correlations over time, which must be rigorously characterised to avoid invalidating upper limits on the 21cm power spectrum . Applying a  filter in fringe-rate space corresponds to convolution of the (transformed) filter in LST space,  so signals will be smeared in time, and noise will become correlated in time. These correlations can subvert common approximations that are used in data analysis pipelines, for instance the way in which the signal-to-noise ratio  is expected to grow as progressively more time samples are averaged together. Some of these issues are addressed in \citet{2018ApJ...868...26C} and \citet{2019ApJ...883..133K}, and analysis needs  to be updated based on the filtering described in this paper. 

The temporal resolution and duration of the observations also limits the resolution of the filter, so that naive Fourier filtering gives rise to unwanted aliasing or ringing effects. We will describe below how this  influenced the implementation of filtering in HERA, based on discrete prolate spheroidal sequences. Finally, flagging due to RFI and other data quality issues  complicates the transformation between the LST and fringe-rate spaces; poor handling of the discontinuities that are introduced by flagging can also lead to ringing artifacts. The use of discrete prolate spheroidal sequences can tolerate some missing data \citep{2021MNRAS.500.5195E}, but we do not examine that in this paper. Inpainting methods such as Gaussian constrained realizations are being considered to deal with flagged data \citep{2023ApJS..266...23K}.

There are other methods that, like fringe-rate filtering, use the   modes present in  time sequences of visibilities to manipulate those visibilities for particular aims. $m$-mode analysis uses time-based spherical harmonic m-modes to decompose the sky and enable foreground subtraction \citep{2014ApJ...781...57S, 2015PhRvD..91h3514S}. Correlations in visibilities over time  can be  used to statistically separate  the EoR signal from foregrounds \citep{2014ApJ...793...28P, 2021MNRAS.504.2062P}. These and other methods provide alternatives to fringe-rate filtering.

In this paper, we apply two different types of fringe-rate filter to simulated  observations. A simulation with a sky model consisting of a single point source and an east-west baseline are  used to build intuition about fringe-rate filtering; these build on the examples  in \citet{2016ApJ...820...51P} but show the use of  discrete prolate spheroidal sequences. Following on from these, full sky simulations using 10 HERA dishes are passed through certain key steps of a `standard' HERA analysis, of the kind used in \citet{2022ApJ...925..221A}. We  examine the outputs of the steps of the analysis, including the final cylindrically-averaged delay spectra, to observe the effect of the filtering.

This paper does not provide a rigorous statistical analysis of  the  effects of fringe-rate filtering and in particular does not address signal loss; these subjects will be dealt with elsewhere \cite[e.g][]{Pascua}.

The paper is organised as follows. Sect.~\ref{sec:frf} provides a brief pedagogical introduction to fringe-rate filtering, using simple examples based on point source sky models to build intuition for how the filters work. Sect.~\ref{sec:sims} summarises the visibility simulations we use in the rest of the paper, while Sect.~\ref{pipeline} defines the mock data analysis pipeline that we apply to the simulations to act as a simplified model of the HERA pipeline. Sect.~\ref{results} presents our main results, which show how the products of each stage of the simplified data analysis pipeline are affected by fringe-rate filtering. Finally, we conclude in Sect.~\ref{sec:conclusions}.

\section{Fringe-rate filtering} \label{sec:frf}

\citet{2016ApJ...820...51P} contains a detailed description of fringe-rate filtering theory and methods, which we will summarise here in the context of our experiments and the implementation of fringe-rate filtering in the HERA pipeline.

\begin{figure*}
    \centering
         \includegraphics[width=0.7\textwidth]{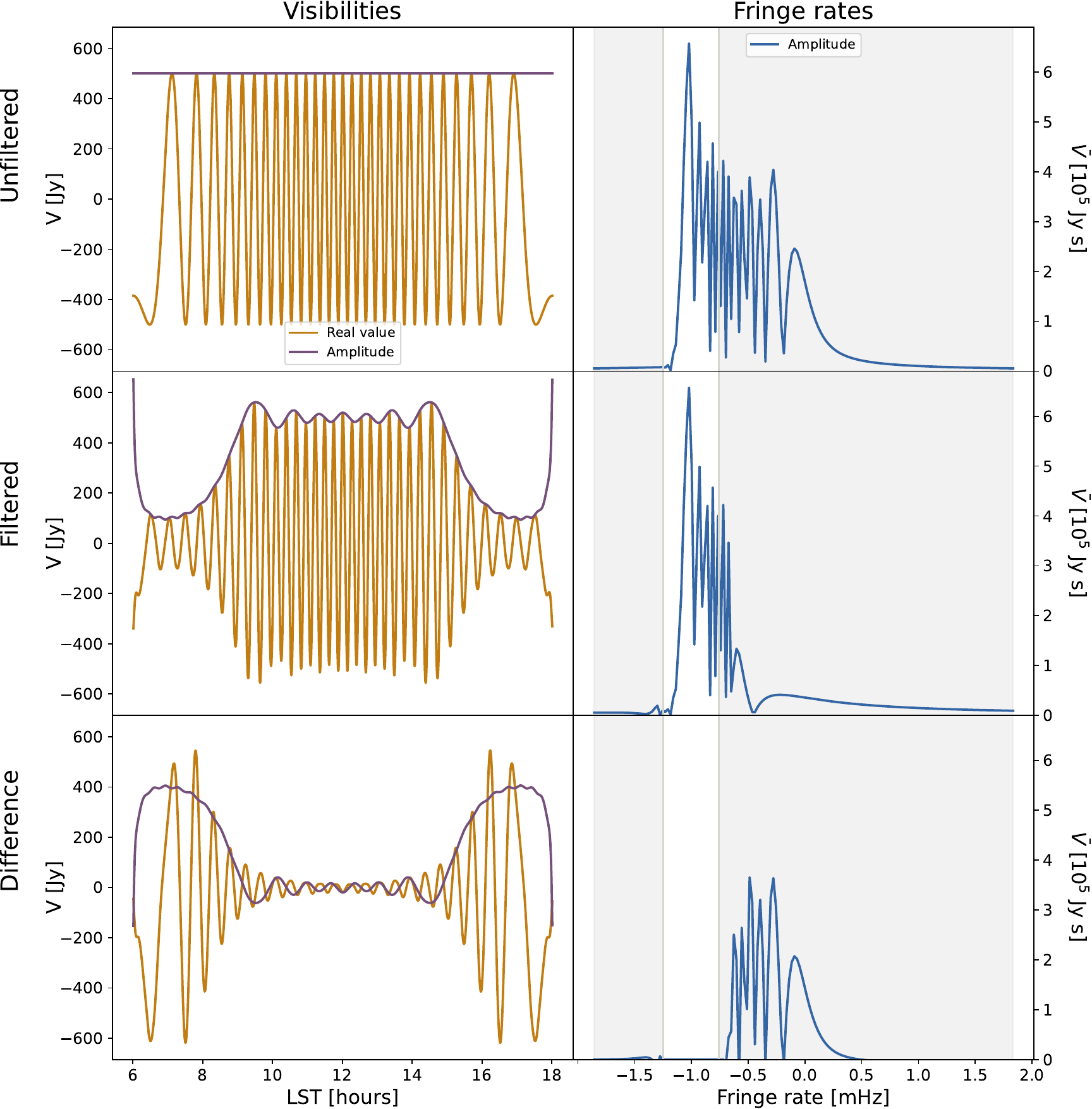}
     \caption{The effect of the mainlobe filter on the visibility for a single source that transits the zenith, as observed by a purely East-West baseline at the equator, with a uniform primary beam. The left column shows the amplitude and real component of the visibility, with fringes clearly indicated by the oscillation of the real component. The right column shows the amplitude of the fringe-rates of the visibilities (the Fourier transform along the time axis). The top row shows the unfiltered visibilities, the second row shows them after filtering with the mainlobe filter, and the third row is the difference between the two. The mainlobe filter specifies that only fringe-rates within the bounds indicated by the grey vertical lines should be retained; the shaded region indicates those fringe-rates that should be excluded.  The filter has been implemented using discrete prolate spheroidal sequences, leading to some leakage outside the bounds, as discussed in the text.}
     \label{mainlobe_ew}
\end{figure*}

A single interferometer baseline correlates the voltage signal received by two antennas, producing a complex visibility for each time sample and frequency channel.
We write the  visibility for a baseline $\boldsymbol{b}$, observing frequency $\nu$, and time $t$,  as
\begin{equation}
    V_{\boldsymbol{b}}(t) = \int d\Omega\, I(\boldsymbol{n}) A({\boldsymbol{n}}, t) \exp \left ( - 2 \pi i \frac{\nu}{c}\boldsymbol{b}(t) \cdot \boldsymbol{n} \right )
    \label{vis_equation}
\end{equation}
where $I$ and $A$ are the sky intensity and primary (power) beam in direction $\boldsymbol{n}$ on the celestial sphere, the integral is over all directions on the sky that are above the local horizon, and we have suppressed the frequency dependence of the terms for brevity. 
For a drift-scan interferometer such as HERA, sources on the sky rotate overhead as the antennas maintain a fixed pointing at the zenith. The received intensity of a source depends on the antennas' directional sensitivity (or `primary beam'), and the phase of the visibility depends on the source direction relative to the baseline. It is the phase that is most important for fringe-rate filtering.

A simplified example of how the phase changes as a source drifts through the primary beam is shown in the top-left panel of Fig.~\ref{mainlobe_ew}. A single source at a declination of 0$^\circ$ is passing over an East-West oriented baseline situated at the equator. The source rises at the East horizon at a local sidereal time (LST) of 6 hours, passes through the zenith at 12~hours, and sets on the West horizon at 18~hours. In this example, we set the antennas to have uniform beams, so that the observed amplitude of the  source is constant over time. The phase, however, changes over time, and is  indicated by showing the real part of the visibility; the imaginary part changes in a similar manner. The oscillations of the phase are the {\it fringes}, and the speed of oscillation is the {\it fringe-rate}. We see from this illustration that the fringe-rate is not constant -- it is lowest when the source is at the horizon, and fastest when the source is at zenith.

If we had performed a Fourier transform along the time axis in a short time window, to a good approximation we would have recovered a single tone, i.e. a spike at a particular fringe-rate, for this source. The specific fringe rate that is observed depends on the baseline length and orientation, but also the time $t$ around which we have placed our short window. Fourier transforming over the whole time range would instead produce a collection of spikes that have been drawn out over a range of fringe-rates due to the variation in fringe rate between horizon and zenith (Fig.~\ref{mainlobe_ew}, top-right panel).

We can now use the varying fringe rate to construct a filter that depends on distance from the zenith. From our example, we can see that the fringe rates with the highest absolute value, at $f \simeq -1$~mHz, correspond to when the source was near zenith. By applying a top-hat filter in Fourier space that select only modes around $f \simeq -1$~mHz (indicated by the grey vertical lines), we can suppress the observed intensity of the source away from zenith, as shown in the middle panels of Fig.~\ref{mainlobe_ew}. The procedure can be generalised to any source declination, array latitude, baseline length and orientation etc., as well as to include the effect of the primary beam, which also modulates the amplitude of sources as they rotate through the beam. In this example, the failure to remove some modes outside the filter bounds (i.e $f \simeq -0.5$~mHz), is due to the use of discrete prolate spheriodal sequences, and will be discussed subsequently.

Following \citet{2016ApJ...820...51P}, we can make some approximations that permit a simplified interpretation of filtering in fringe-rate space. By relating the change in baseline orientation with time to the rotation of the Earth, and making an approximation that the fringe pattern varies faster with time than the primary beam term, we can replace the primary beam term in Eq.~\ref{vis_equation} with an effective `filtered' beam \citep{2016ApJ...820...51P},
\begin{equation}
    V^{\rm filt.}_{\boldsymbol{b}}(t) = \int d\Omega\, I(\boldsymbol{n}) A^{\rm filt.}_{\boldsymbol{b}}({\boldsymbol{n}}, t) \exp \left ( - 2 \pi i \frac{\nu}{c}\boldsymbol{b}(t) \cdot \boldsymbol{n} \right ),
\end{equation}
where
\begin{align}
    A^{\rm filt.}_{\boldsymbol{b}}(\boldsymbol{n}, t) &\approx A_{\boldsymbol{b}}(\boldsymbol{n}, t) \int df w(f)\, \tilde{\gamma} \left ( \frac{\nu}{c}(\boldsymbol{b}(t) \times \omega_\oplus) \cdot \boldsymbol{n} - f \right ). \label{fringe_rate_equation}
\end{align}
Here, $w(f)$ is a weighting in fringe-rate space corresponding to the chosen fringe-rate filter, $\tilde{\gamma}$ is the inverse Fourier transform of a tapering function that determines how long the time window used for the fringe-rate transform is, and $\omega_\oplus$ is the angular velocity vector corresponding to the Earth's rotation. If the time window $\gamma$ is broad, $\tilde{\gamma}$ is highly peaked in the fringe-rate domain, and so the convolution is localised around $f \approx (\nu/c)(\boldsymbol{b} \times \omega_\oplus) \cdot \boldsymbol{n}$. For a given baseline vector $\boldsymbol{b}$, this determines a fringe-rate for each direction on the sky, $\boldsymbol{n}$. Fringe-rates are not unique to each direction however; instead, there are regions of constant fringe-rate which form rings on the sky \citep{2016ApJ...820...51P}. These can be mapped more or less crudely to different regions of the primary beam, hence the interpretation of fringe-rate filtering as beam sculpting.

Note that some baseline orientations (i.e. mostly North-South aligned ones) have $\boldsymbol{b}(t) \times \omega_\oplus \approx 0$. For these, most regions on the sky are collapsed into a narrow range around $f = 0$~mHz, making fringe-rate filtering impractical for these baselines.

\subsection{Fringe-rate filters for HERA}

In this work, we will focus on two types of fringe-rate filter that will be used in forthcoming HERA analyses. The first is the `mainlobe' filter, which implements a tophat window in fringe-rate space that rejects components of the observed visibilities with fringe rates that fall outside the mainlobe of the primary beam. This down-weights the primary beam sidelobes and horizon, which suffer most from chromatic effects that are an important source of systematic contamination, particularly when bright sources such as Fornax~A pass through them. By removing these fringe-rates, we can hope to reduce the amount of data that has to be flagged due to systematic contamination. This filter can be thought of as a very basic form of beam sculpting. We do not attempt to re-weight the primary beam by applying a shaped filter however; it is purely a cut on the range of fringe-rates to retain only ones that are within the beam mainlobe. 

\begin{figure*}
    \centering
         \includegraphics[width=0.7\textwidth]{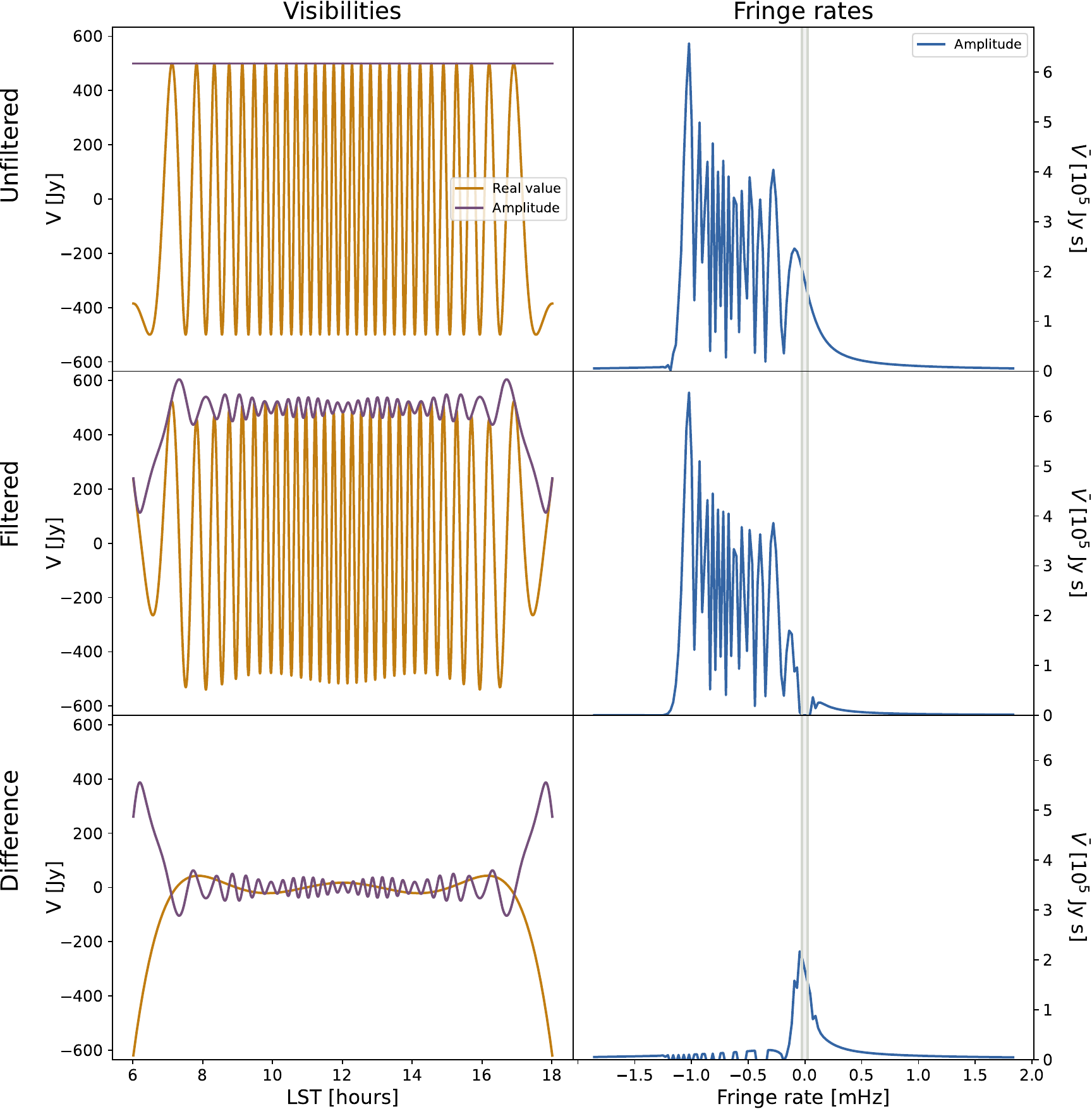}
     \caption{Similar to Fig.~\ref{mainlobe_ew} but now for the notch filter. The notch filter specifies that fringe-rates around $f = 0$~mHz should be excluded, indicated by the shaded region. The filter was implemented using discrete prolate spheroidal sequences, as discussed in the text.}
     \label{notch_ew}
\end{figure*}

The second is a `notch' filter, which is intended to remove a narrow range of fringe rates around $f \approx 0$~mHz. These correspond to components of the signal that are not fringing, i.e. those that are stationary with the array. While some of the sky signal can fall within this range (e.g. near the celestial poles), these fringe rates are also contaminated by systematic effects like mutual coupling \citep{2019ApJ...884..105K}.


The filters are specified as rectangular tophats in fringe-rate, with centre $f_c$ and half-width $\Delta f$. For the mainlobe filter this specification defines the fringe-rates that are to be kept, and the filters vary with baseline and frequency. For the notch filter, this defines the fringe-rates to be rejected, typically a small range around 0~mHz, and the filter specification is the same for all baselines.

In the ideal case, the two tophat filters could be implemented by Fourier transforming the visibilities along the time dimension, masking in/out the appropriate range of fringe-rates, then and transforming back. Several complications arise for real data which means that a more sophisticated implementation must be used however.

Of most concern is missing data. For any given set of observations, a substantial fraction of times will be flagged due to RFI, poor calibration solutions, or system problems. The missing data appear as sharp features to a Fourier transform, and so generate ringing that scatters parts of the signal across fringe-rate space. A related phenomenon is due to the non-periodicity of the data, which is seen by the Fourier transform as another discontinuity at the band edges.

To reduce ringing artifacts, a tapering function such as a Blackman-Harris window \citep{1978IEEEP..66...51H} can be applied to the visibility data to enforce periodicity or apodise the edges of the flagged regions. Alternatively, more sophisticated spectral analysis methods can be used that avoid direct Fourier transforms and instead use basis functions with properties that reduce the risk of ringing. HERA uses the latter kind of filtering method, based on discrete prolate spheroidal sequences \citep[DPSS;][]{mathews, moore, 2021MNRAS.500.5195E}. These are a set of orthonormal basis functions that are  concentrated in Fourier space for a given tophat window. In other words, they are constructed to minimise  the total amount of power that is able to leak outside the tophat region in the fringe-rate domain. 

However, DPSS are not all created equal, in terms of how well they concentrate power. For this reason it is possible to create a set of DPSS  that  are a complete basis for a sequence of visibilities; the sequence can then be exactly decomposed into DPSS, much like a Fourier transform for Fourier modes.  For filtering, we generally only want to retain a few DPSS that are most concentrated within the filter window. The more DPSS that are retained, the better the approximation to a tophat filter in fringe-rate with sharp edges, but the worse the leakage outside the tophat window region. The HERA filtering pipeline has several options available to choose the number of DPSS to retain in an automated fashion, e.g. based on a minimum permitted concentration ratio \citep{2021MNRAS.500.5195E}. 

Once we have the reduced set of DPSS, we  transform each of them back into the LST domain and use a linear least-squares fitting algorithm to find their best-fitting coefficients. This approach naturally handles missing data and non-periodicity at the band edges. The filter is  only an approximation to a tophat, and will have some degree of leakage outside the tophat region and slight suppression of power inside the region. For clarity, note that a taper is not applied in the time dimension when performing the DPSS transform, and so some edge effects can still be present.


Because the filter is implemented through this fitting procedure, there is no simple, unique, filter profile that can be plotted. We will, however, present examples of the filter in action, which illustrate a typical filter shape when comparing  unfiltered and filtered data.

Finally, we note that the shape of the filter is also limited by the resolution in the fringe-rate domain, which is determined by the length of the time/observing window, $\Delta t$. Longer time windows permit higher resolution, i.e. $\delta f = 1/\Delta t \approx 0.05$~mHz for $\Delta t = 6$~hours. 

The advantages and disadvantages of DPSS for filtering 21cm observations is discussed more fully in \citep{2021MNRAS.500.5195E}.
A simple  example of fitting DPSS, for the mainlobe filter, is given in Appendix \ref{app1}. We use the eigenvalue cutoff method for selecting highly concentrated DPSS; its meaning, the value we used, and its consequences, are explored in Appendix \ref{app2}.

\subsection{Examples of filters applied to idealised situations}

At the beginning of Sect.~\ref{sec:frf} (and shown in Fig.~\ref{mainlobe_ew}), we discussed the effect of a mainlobe filter on the highly simplified situation of a single point source transiting through the zenith as observed by an East-West baseline at the equator with a uniform primary beam. Returning to this example for a more detailed examination, we see that most of the signal lies between fringe-rates $-2.2 \lesssim f \lesssim 0.5$~mHz. The predominantly negative rate is due to the direction of Earth's rotation. The few positive fringe rates in this example are an artifact of the Fourier transform to fringe-rate domain, as we have not used a taper to enforce periodicity, and the uniform primary beam also fails to provide any tapering of its own. In general, some positive fringe-rates are expected to be observed from sky emission however, corresponding to sources on the far side of the celestial pole relative to the beam centre.

The determination of the fringe-rate limits is based on beam power, which is low towards the horizon and highest at zenith. These also correspond to locations where the fringe-rates are low and high, respectively. The  fringe-rates are plotted against the beam power, and cut based on the 5\% and 95\% percentile of the beam power. The low and high fringe-rates become the filter limits. The high limit of 95\% is used simply based on the evidence of Fig. \ref{mainlobe_ew}, where it cuts out the low fringe-rates after the peak; the behaviour of the filter on the baselines of an actual telescope should be examined to see if this is reasonable. Other schemes are also possible, such as basing the low fringe-rate limit on beam nulls, but these have not yet been implemented. The same scheme and limits are used for all the mainlobe filtering experiments in this paper.

To complement this example, we also plot the effect of the notch filter on the same scenario in Fig.~\ref{notch_ew}. The mainlobe filter retains fringe-rates near the maximum rate, and the notch filter rejects fringe-rates near $f \simeq 0$~mHz, but the filtered data has features in common, for our simple examples. For the notch  filter, the  amplitude of the visibilities has been reduced closer to the horizon (LSTs of 6~hours and 18~hours) where the fringe rate tends to 0~mHz. The amplitude suppression kicks in much closer to the horizon than for the mainlobe filter however, which is to be expected as the notch filter is much narrower and therefore has an effect that is more localised on the sky. The notch filtered signal in the fringe-rate domain (second row, Fig.~\ref{notch_ew}) shows that a clear notch has been cut out around 0~mHz, but the notch feature does not have a sharp cutoff due to the use of the DPSS fitting method described above. This is analogous to the leakage outside the filter bounds in the mainlobe filter.

Despite using the DPSS fits, additional oscillatory structure has still been imposed on the fringe amplitudes after filtering, which can be seen clearly in the amplitude difference plots (bottom row). The bottom-right panel  of Fig.~\ref{notch_ew}, for example, shows the source of some of this structure -- there is a low-level ripple at negative fringe-rates far outside the notch, and a tail at positive fringe-rates. These tails are an artifact of the DPSS fits, which trade-off filter sharpness and accuracy of signal recovery (i.e. lack of signal loss) inside the filter region against the level of leakage outside the filter region. This trade-off is controlled by the number of DPSS that are retained in the  filter, which can be tuned depending on what one is trying to achieve with the filter.

We do not show the effect of the filters on a North-South baseline because they can be described simply. For a  North-South baseline at the equator, with a source passing over the equator, their is no fringing, i.e. the fringe-rate is 0~mHz. The notch filter removes fringe-rate 0~mHz and so removes all the visibilities, setting the baseline to 0 over all time. Sky locations cannot be identified by fringe-rate for this baseline, so  the mainlobe filter has filter bounds centered at 0~mHz and a half-width set by the fringe-rate resolution (or some other method); it passes all the visibilities without alteration.


Having gained some intuition as to the use and effect of filters in simple situations, we now turn to an examination of the filters that are being used in the HERA pipeline, and how these affect the 2D power spectra of more realistic visibility simulations.

\begin{figure}
\centering
\includegraphics[width=\columnwidth]{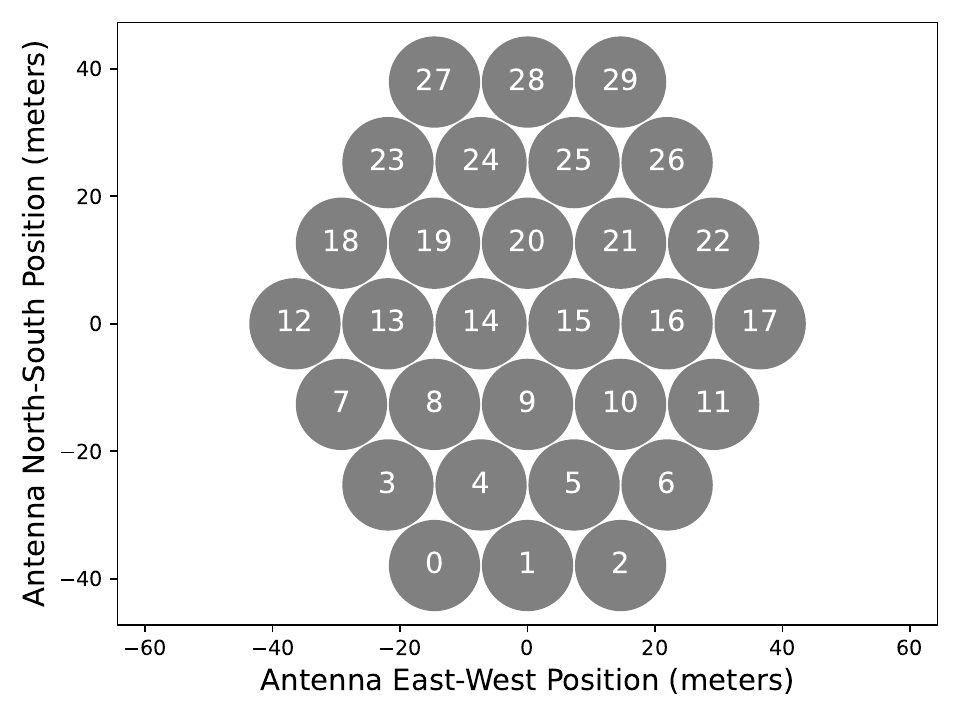}
\includegraphics[width=\columnwidth]{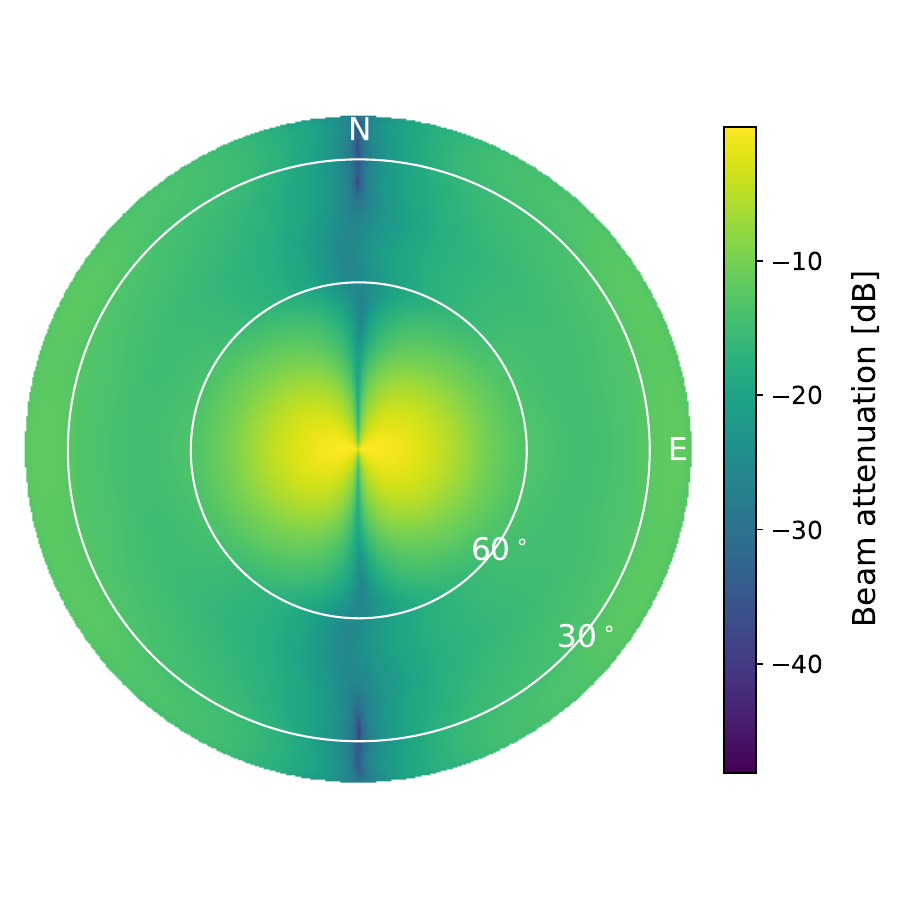}
\includegraphics[width=\columnwidth]{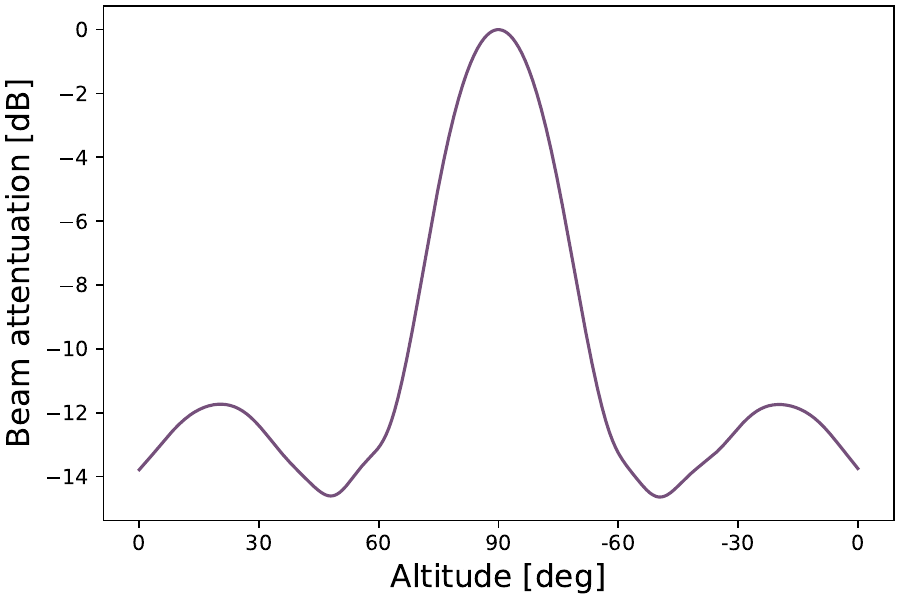}
\caption{{\it (Top):} The antenna layout of the HERA-like array of 30 dishes used for the simulations. Individual antennas are numbered for reference. The minimum spacing is 14.6m. {\it (Middle):} The model HERA E-field beam (see text) for polarization X, 100 MHz,  with East (E) and North (N) marked. The beam amplitude is expressed in decibels relative to a peak of 0 dB.  {\it (Bottom):} A slice through the model HERA  beam   from East to West as a function of altitude. }
\label{fig:layout}
\end{figure}

\begin{table}
\centering
\begin{tabular}{|c|c|}
\hline
{\bf Property} &  {\bf Value} \\
\hline
Antennas & 30 \\
Baselines & 465 \\
\hline
LST range & 7.44 -- 13.46~hours  \\
Integration time & 10 sec \\
Time samples & 2160 \\
\hline
Frequency range & 100 -- 140 MHz \\
Frequency channels & 408 \\
Channel width & 97 kHz \\
\hline
Point sources &20,000 sources (GLEAM-like) \\
Diffuse emission & PyGSM (HEALPix \texttt{NSIDE = 128}) \\
\hline
\end{tabular}
\caption{Summary of the visibility simulation parameters.}
\label{sim_params}
\end{table}

\section{Visibility simulations} \label{sec:sims}

In this section, we describe the simulations that will be used in the remainder of the paper. These are intended to be reasonably realistic as far as the visibilities are concerned; we include a full sky model containing point sources and diffuse emission, a realistic HERA beam model, and simulate a set of baselines for a HERA-like array using 30 dishes in the same location as the HERA telescope. The simulations cover a broad range and many samples in time and frequency to permit high resolution in the spectral (delay/fringe rate) domains. For simplicity, we do not include flagging, bandpass or gain calibration errors, or non-redundancy of the antennas or baselines, nor do we attempt to model the systematic effects that are one of the principal reasons for implementing fringe-rate filtering. The parameters of the simulations are summarised in Table~\ref{sim_params}.

Simulated visibilities are generated using the \texttt{hera\_sim} and \texttt{matvis} packages \citep{2023arXiv231209763K}. We use a catalogue of 20,000 point sources that mimic the distribution of sources in the GLEAM catalog \citep{2017MNRAS.464.1146H}, but fill-in areas of the sky that GLEAM does not cover \citep[for more details, see][]{2021MNRAS.506.2066C}. The point source flux densities for this catalogue range from $0.2 - 82.4$~Jy, with spectral indices from $-1.01$ to $-0.63$. A diffuse emission model is obtained from the PyGSM package \citep{2016ascl.soft03013P}, with the emission represented as a HEALPix map per frequency, with \texttt{NSIDE = 128}. This results in $\sim 27^\prime$ pixels, which are sufficiently small compared to the characteristic angular scale of the longest baseline ($\lambda / D \approx 96^\prime$ for the 77m baseline at 140~MHz) that we expect discretisation effects to be sub-dominant in our results. As an approximation, the visibility simulators treat the centre of each pixel effectively as a point source. We do not include a 21cm component in the simulations as it is much fainter than the foreground components, and we do not study the effects of the filters on it in the present work.


The subset of 30 HERA dishes with its close-packed hexagonal layout shown in Fig.~\ref{fig:layout} (Top). The dish separation is 14.6~m (the shortest baseline length), with a longest baseline of 73~m. We label baselines using their constituent antennas, e.g. baseline (3, 6) runs East-West and has length 48.3~m. The (E-field) primary beam for each antenna is a frequency-dependent dipole-like model derived from electromagnetic simulations described in \citet{2021ITAP...69.8143F} (the ``Vivaldi'' beam).  The beam for $X$ polarization, at 100~MHz, is shown in Fig. \ref{fig:layout} (Middle). A slice from East to West gives the profile in Fig. \ref{fig:layout} (Bottom). We generate only the $X$ (East-West aligned) polarization channel in the simulations.

To permit high-resolution studies of the filters, we simulate large ranges in frequency and time. The time axis comprises 6 hours of HERA drift-scan observations, with 10s of integration time per time sample, starting from Julian date 2458116.48, for a total of 2160 time samples. This spans a range in local sidereal time (LST) of 7.44 -- 13.46 hours. 
The frequency range is $100 - 140$~MHz, comprised of 408 channels of width 97 kHz. Using these frequencies, the 21cm line would be observed over a redshift range of $ z \approx 9.1-13.2$; real-world analyses tend to use smaller bands for power spectrum estimation however, to reduce the effects of cosmic evolution.


After simulating visibilities for the sky model, uncorrelated white noise is added, using the radiometer equation with $T_{\rm sys} \approx T_{\rm sky}$ to set the noise level \citep[see][]{2021MNRAS.506.2066C}.

\begin{figure*}
\centering
\includegraphics[width=\textwidth]{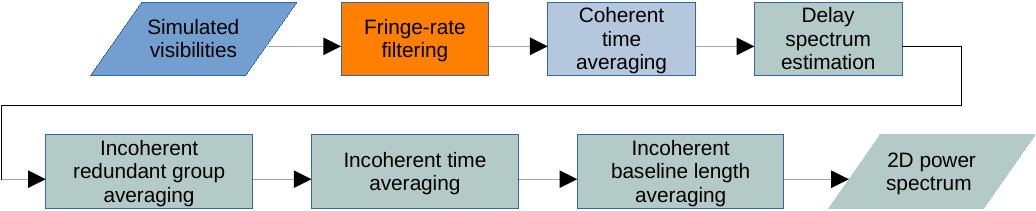}
\caption{Schematic of the data processing pipeline used for this work, showing the steps taken to produce a 2D power spectra, and the placement of the fringe-rate filtering step.}
\label{fig:flowchart}
\end{figure*}
\section{Data processing pipeline} \label{pipeline}

In this section, we describe the analysis pipeline that we apply to the simulated visibilities (we do not perform imaging). This is an idealised, cut-down version of the HERA pipeline that was used in \citet{2022ApJ...925..221A} and \citet{2022ApJ...924...85A}. A flowchart summarising the steps of the pipeline is shown in Fig.~\ref{fig:flowchart}.

The purpose of the simulated pipeline is to apply a given fringe-rate filter to the input visibilities and then form per-baseline delay power spectra, followed by a sequence of averaging and binning operations to increase the signal-to-noise. The end result is a 2D power spectrum indexed by baseline length and delay which can be converted to cosmological wave modes $k_\perp$ and   $k_\parallel$, although we retain baseline length ($|b|$) and delay ($\tau$) for plots in this paper. The 2D power spectrum can be further averaged into a spherical power spectrum if desired. 

Simulated visibilities are fringe-rate filtered before being coherently averaged in time, in chunks of 300~s. The visibilities are re-phased to zenith before averaging the chunks, in order to prevent time smearing, and then re-phased back to a drift-scan observation. This coherent averaging step is designed to average down the noise while causing minimal signal loss, as described in \citet{2022ApJ...924...85A}. The optimal chunk size for coherent averaging is baseline- and frequency-dependent, and should generally be significantly shorter than the beam-crossing timescale (the time taken for a patch of sky to drift through the mainlobe of the primary beam). Chunk lengths of around 420~sec have been proposed for HERA \citep{2022ApJ...924...85A, 2022ApJ...925..221A}, but we made a more conservative choice of 300~sec for this paper.

The pipeline is run twice, once with the mainlobe filter and once with the notch filter. In the finalized HERA pipeline, it is unlikely that the filters will be separated in this way, they will both appear in the pipeline at different stages before coherent averaging (for example separated by an LST binning step when multiple days are used). However, so that we can examine them separately, we run them separately.

Following coherent time averaging, pairs of baselines that are redundant with each other (having the same length and orientation) are Fourier (delay) transformed and multiplied to form delay power spectra. This step is implemented using an optimal quadratic estimator  \cite[see][]{2022ApJ...925..221A, 2014PhRvD..90b3018L}, but with a uniform weighting, which is not optimal. A Blackman-Harris taper \citep{1978IEEEP..66...51H} is used for the Fourier transform. Pairs consisting of the same baseline (`auto-baseline pairs') are excluded, as these would be affected by a noise bias. After coherent averaging, delay spectra that are redundant with each other are averaged together. Due to the large size of the data and computing constraints, we limit the sets of redundant delay spectra to be  averaged to a maximum of 48 delay spectra. Then follows 
time-averaging followed by binning by baseline length.

Note that we run the pipeline on the simulated data both with (`filtered') and without (`unfiltered') the fringe-rate filtering step, so we can study the effect of the different filters by comparing unfiltered vs. filtered. As we are also interested in the effect of the filters on the noise, and how this affects the final signal-to-noise of the averaged data, we also run the pipeline on simulated visibilities without noise added (`noiseless'), so the results can be compared with the `noisy' data.

\begin{figure*}
\centering
\includegraphics[width=0.8\textwidth]{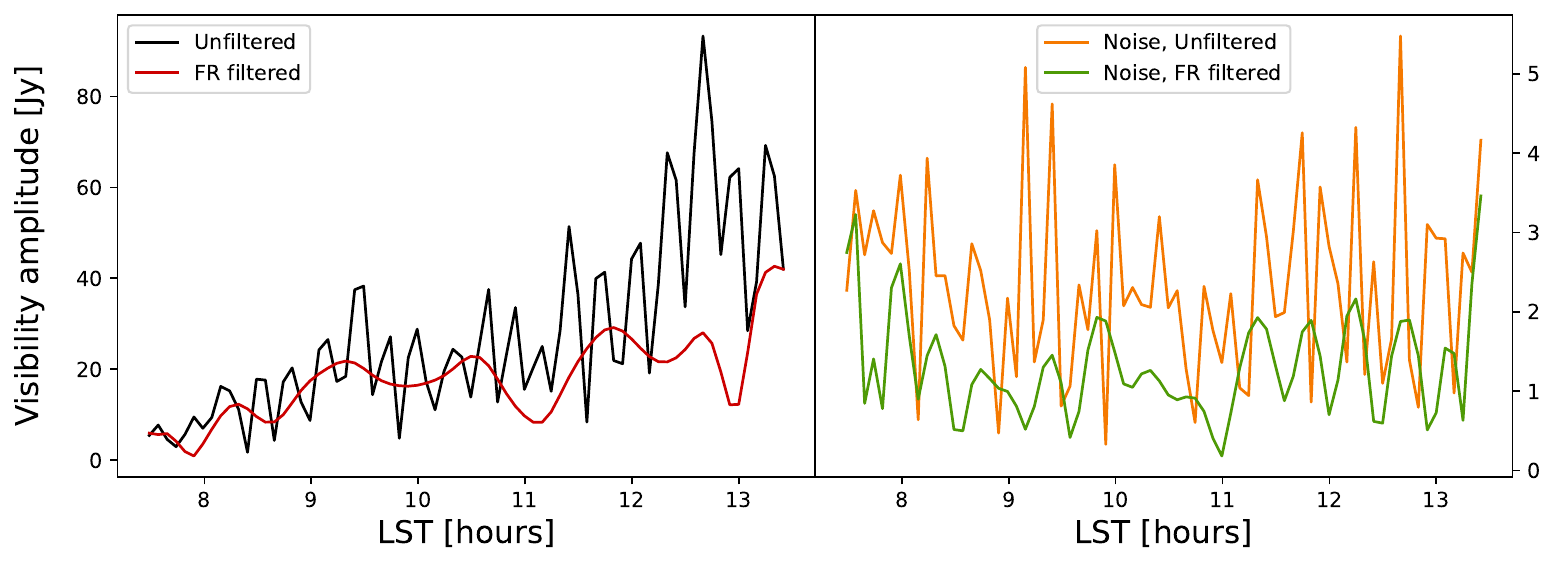}
\includegraphics[width=0.8\textwidth]{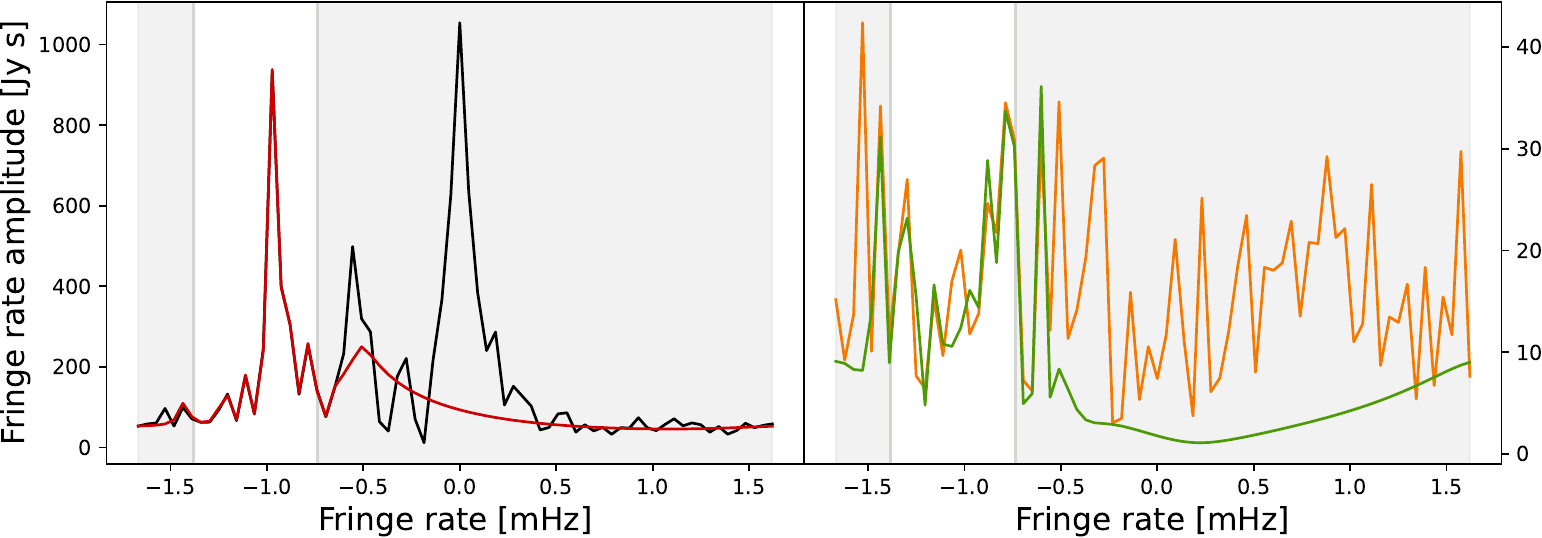}
\includegraphics[width=0.8\textwidth]{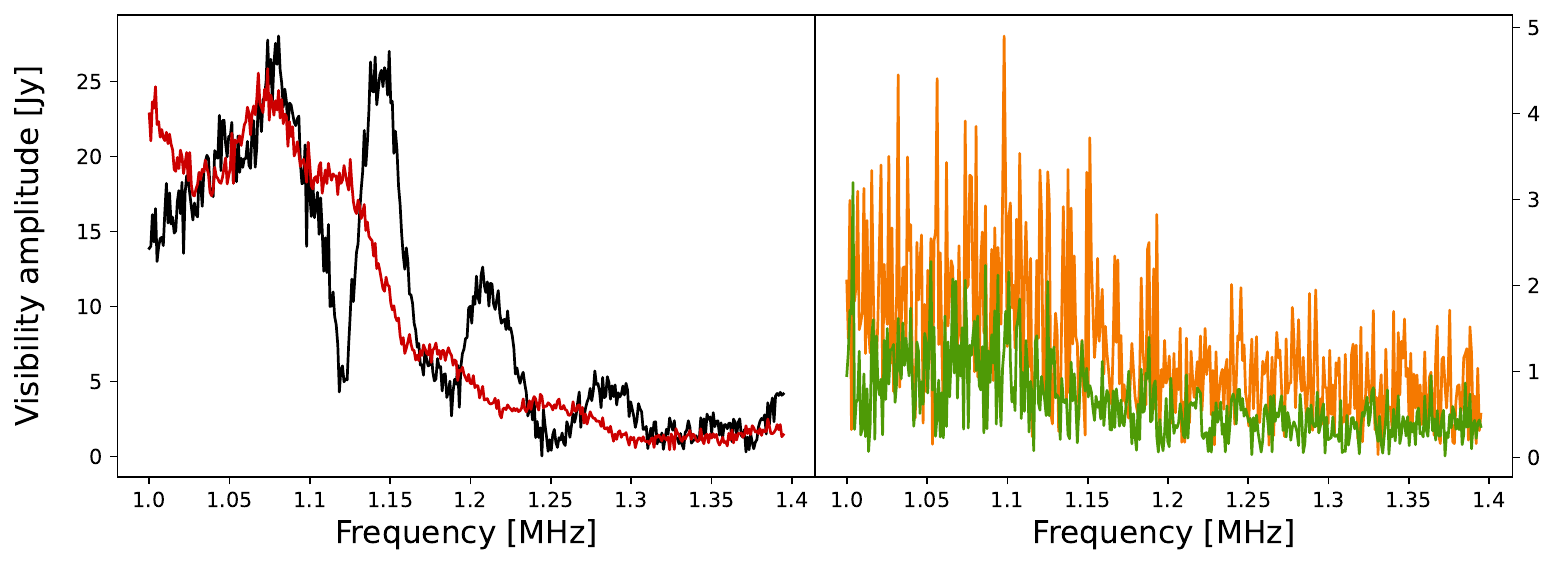}
\includegraphics[width=0.8\textwidth]{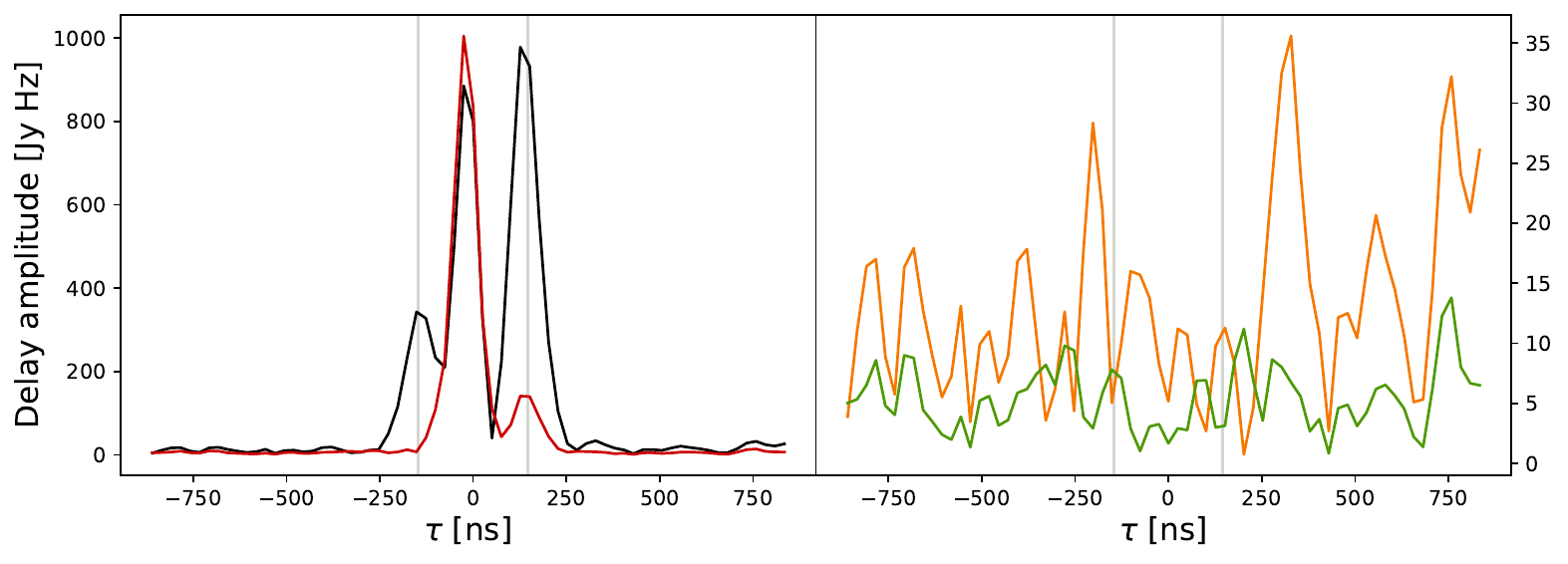}
\caption{The visibilities and fringe-rates for baseline (3,6)  before and after application of the mainlobe filter. The top row left column shows the amplitude of visibilities over the time range of the simulation, both unfiltered and fringe-rate (FR) filtered using the mainlobe filter, at a frequency of 100~MHz.
 The top row right column shows the noise in the visibilities. The second row shows the  fringe-rate transform of those visibilities, on the left, and the noise in the fringe-rate transform on the right. The shaded region indicates the fringe-rates that should be excluded by the mainlobe filter. The third row left column shows
 the amplitude of visibilities over the frequency range of the simulation, at the midpoint of the simulation time range (an LST of 10.45~hours), again showing visibilities on the left and the noise in the visibilities on the right. The fourth row shows the delay transform obtained from the third-row visibilities,
 and the noise in the delay transform. The grey lines show the horizon delay for this baseline.}
\label{mainlobe_vis_summary}
\end{figure*}

\begin{figure*}
\centering
\includegraphics[width=0.8\textwidth]{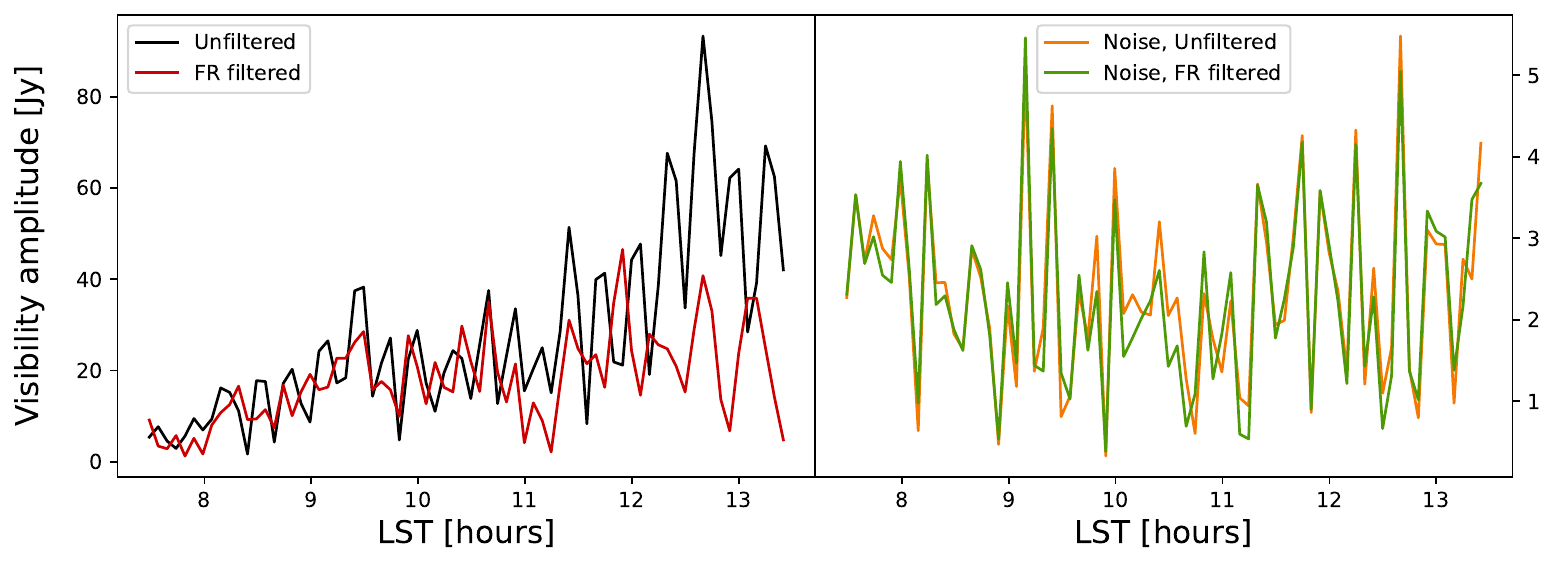}
\includegraphics[width=0.8\textwidth]{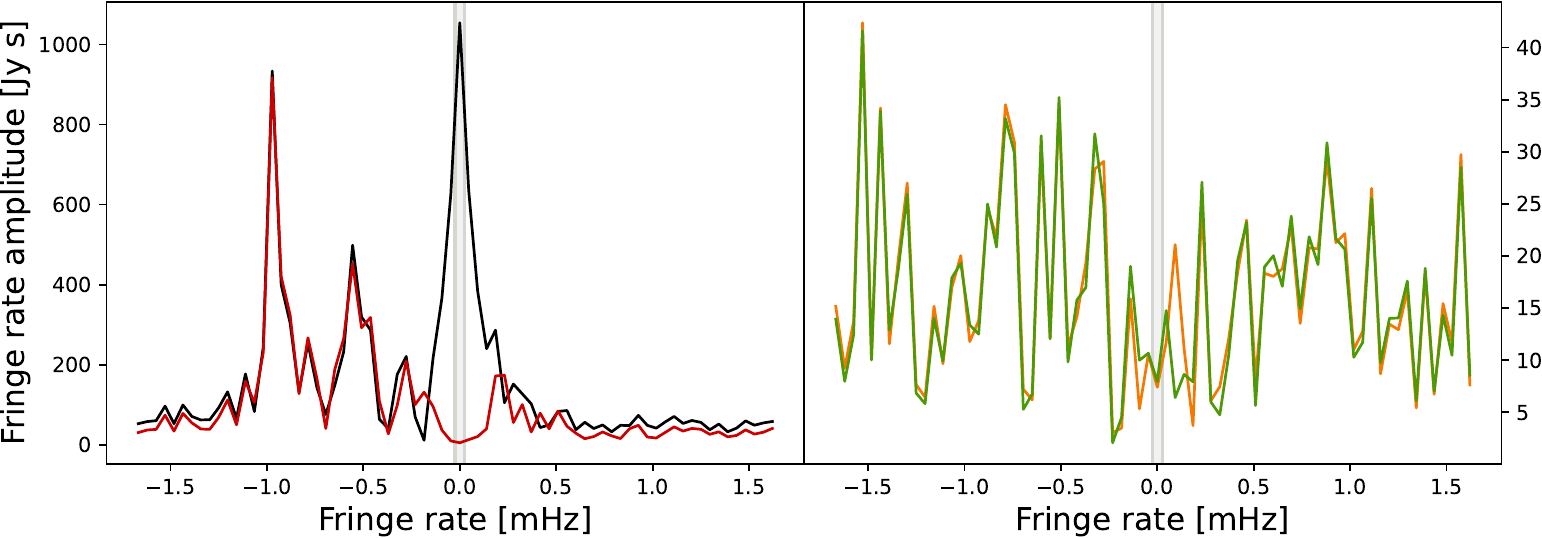}
\includegraphics[width=0.8\textwidth]{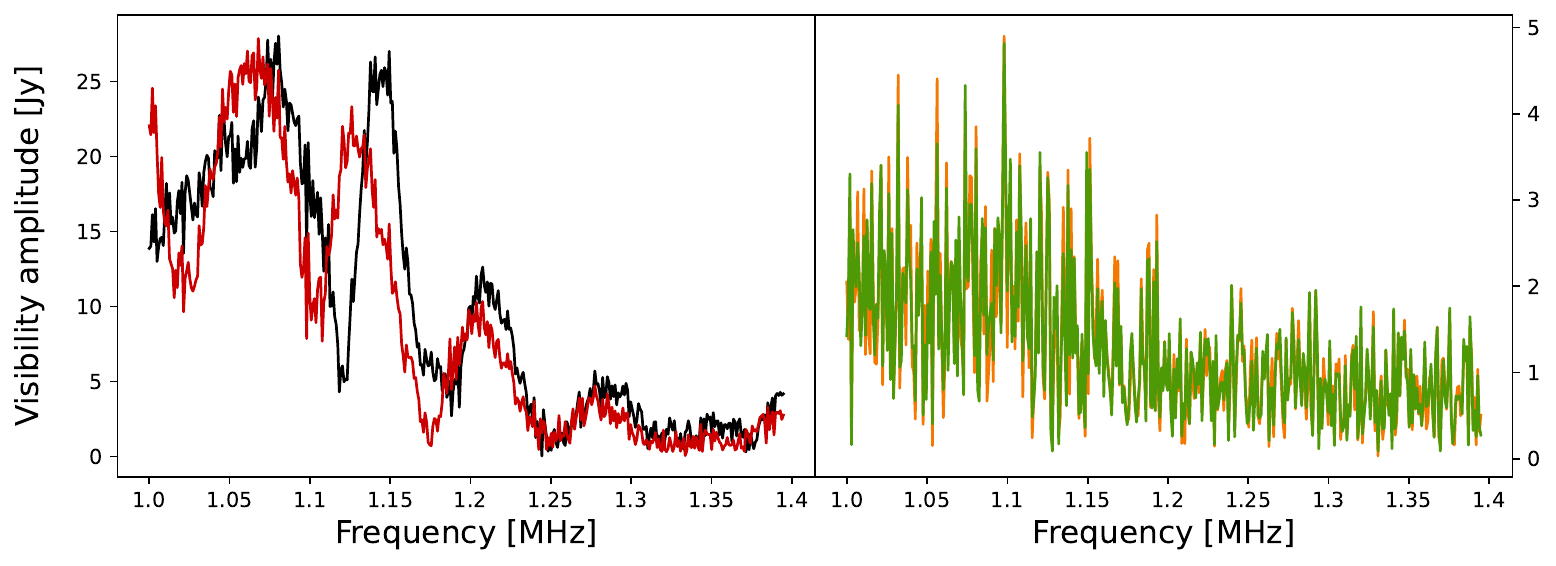}
\includegraphics[width=0.8\textwidth]{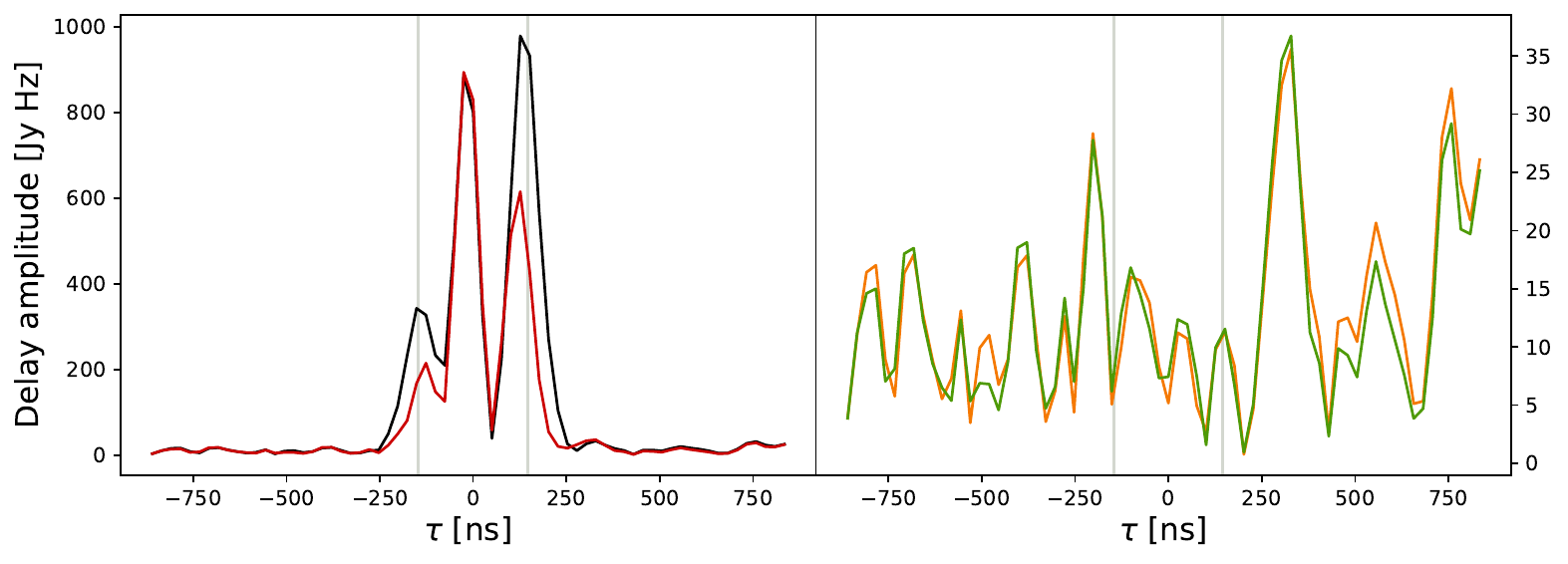}
\caption{As for Fig.~\ref{mainlobe_vis_summary}, but the notch filter is used.}
\label{notch_vis_summary}
\end{figure*}

\begin{table}
\centering
\begin{tabular}{|c|c|c|c|c|c|c|}
 {\bf Visibility axis} & \multicolumn{2}{c}{\bf Unfiltered} & \multicolumn{2}{c}{\bf Mainlobe} & \multicolumn{2}{c}{\bf Notch} \\
 \hline
  &  Mean &  Std. & Mean &  Std. & Mean & Std. \\
 \hline
 {\bf Time} [Jy]    & 0.05  & 1.85  & $-$0.04  & 0.96 & 0.05  & 1.82 \\
 {\bf Freq.} [Jy] & & $-$3.58  & 13.22  & $-$3.49  & 8.41 & $-$1.99   \\
 \hline
 {\bf Fr.-rate}  [Jy\,s]             & $-$0.04  & 1.16  & 0.00  & 0.55 & $-$0.03  & 1.13  \\
 {\bf Delay}        [Jy\,Hz]            & $-$0.69  & 21.62  & 0.85  & 11.37 & $-$0.68  & 21.50   \\
 \hline
 \end{tabular}
 \caption{Statistics of the real component of the noise for different axes of the visibility data, as plotted in the right columns of Figs.~\ref{mainlobe_vis_summary} and  \ref{notch_vis_summary}. The unfiltered data is the same in both figures. The rows are not directly comparable, as they have different units or correspond to different fixed times/frequencies. }
\label{vis_noise_statistics}
\end{table}

\section{Results} \label{results}

In this section, we present the results of applying the mainlobe and notch filters to the simulated data, examining the data at different points in the mock analysis pipeline. We begin by examining the effect on the visibilities themselves (Sect.~\ref{sec:res:vis}), then examine different stages in the delay spectrum averaging process (Sect.~\ref{sec:res:dspec}), followed by the final 2D power spectrum results (Sect.~\ref{sec:res:2dspec}). 


\subsection{Effect of filters on visibilities} \label{sec:res:vis}

In this section we compare the filtered and unfiltered visibilities obtained from  the output of  the coherent time averaging step, which uses a 300~sec averaging window and therefore reduces the number of time samples to 72 from the original 2160.

Fig.~\ref{mainlobe_vis_summary} shows the visibilities (left column) and noise (right column) for baseline (3, 6) (in Fig. \ref{fig:layout}, top), which is East-West orientated and has length 43.8m. The top row shows the visibilities over time at a fixed frequency of 100~MHz before and after the mainlobe filter has been applied. The second row shows the same data in fringe-rate space (i.e. after applying a Fourier transform in the time dimension, without a taper function). The third row shows the visibilities as a function of frequency at a fixed LST of 10.45~hours, and the fourth row shows the same data in delay-space, that is, after Fourier transforming the visibilities by frequency (using a Blackman-Harris taper),  at fixed LST. We have restricted the delays to focus on the horizon (grey vertical lines); the full delay range for our simulations is $-5154$ to $5129$ ns.

To obtain the noise plots, we processed the noisy and noiseless visibilities identically and took the difference.

We see that the visibilities are smoother in time than in frequency due to the coherent time averaging step. The mainlobe filter causes substantial additional smoothing in time, which has also reduced the noise level, as anticipated. The visibility as a function of frequency (at fixed time) shows a significantly different structure following filtering, which now lacks some of the significant structures on scales of a few MHz which are seen in the unfiltered data, while the larger-scale and smaller-scale structures are less affected. 
The delay-space plot shows more clearly how the delay structure of the visibility (at fixed time) has changed as a result of the mainlobe filter, with a notable suppression of power around the horizon delays. This is the expected behaviour for a mainlobe filter; the sidelobes and horizon, which are also more chromatic and therefore appear at higher delay, have been suppressed.

The fringe-rate plots show that the fringe-rates inside the tophat are recovered with hardly any loss of signal, while some fringe-rates outside the filter bounds have survived, but at a significantly reduced level. In fact, the filtered signal at fringe-rates $f \gtrsim -0.6$~mHz is mostly caused by leakage of the signal within the tophat bounds outside the filter region due to the DPSS fitting procedure, rather than being a residual of the original unfiltered signal at these fringe-rates. It can be seen that the structure of this leakage signal is quite different as a function of $f$, compared  to the unfiltered signal. The large peak around $f \approx 0$~mHz has been removed however, and the smaller one at $f \approx -0.6$~mHz has also been significantly reduced. At $f \gtrsim 0.5$~mHz, the filtered signal level is sometimes slightly above the unfiltered signal level however, again due to the DPSS leakage. As discussed above, the level of leakage can be tuned by changing the number of DPSS  that are retained in the fitting procedure, at the expense of making the tophat filter boundary less sharp, and potentially increasing the level of signal loss inside the tophat region. The noise in the fringe-rate domain has also been strongly suppressed outside the tophat region, but left more or less intact inside the tophat and immediately around its boundary.

Fig.~\ref{notch_vis_summary} shows the same quantities as Fig.~\ref{mainlobe_vis_summary}, but now for the notch filter. This time, the main effect of the filter has been to remove the large spike around $f \approx 0$~mHz. There is some minor signal loss outside of the filter range, particularly at large $|f|$, and as expected for the DPSS fit, the shape of the filter is not perfectly sharp. The noise as a function of time and frequency is broadly similar before and after filtering, reflecting the narrowness of the notch filter, which can only remove a small fraction of the noise. Both the time and frequency structure of the visibilities is quite similar before and after the filtering, with only small shifts in the spectral features and a small reduction in signal amplitude observed. 


Table \ref{vis_noise_statistics} gives means and standard deviations for the real component of the noise in Figs. \ref{mainlobe_vis_summary} and \ref{notch_vis_summary}. These show that the mainlobe consistently  reduces the noise by almost one half, whereas the notch filter has little effect, as can be seen from the plots.



\begin{figure*}
\centering
\includegraphics[width=0.95\textwidth]{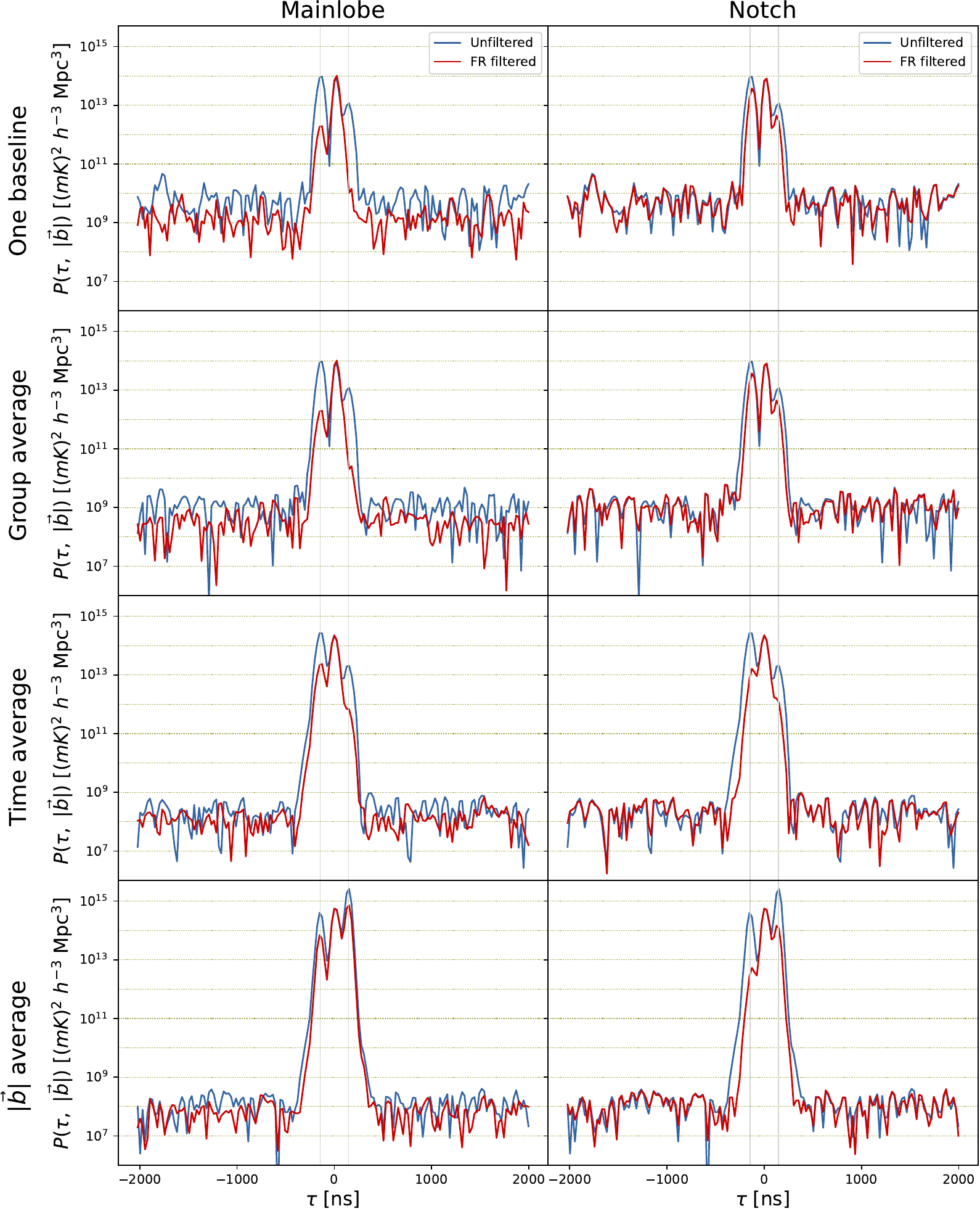}
\caption{Shows the delay spectra for baseline-pair (3,6) and (8, 11) as it proceeds through the pipeline, being combined with  delay spectra from redundant baseline pairs (Group average), averaged over time (Time average), and combined with other delay spectra of the same baseline length (|$\vec{b}$| average). The left column shows delay spectra generated using the mainlobe filter, the right column shows delay spectra generated using the notch filter. The vertical grey lines indicate the horizon delay. The stages of the pipeline are listed on the left.}
\label{delay_spec_summary_both}
\end{figure*}

\subsection{Effect of filters on averaged delay spectra} \label{sec:res:dspec}


HERA uses a delay power spectrum approach \citep{2012ApJ...756..165P}. Visibilities are Fourier transformed to delay ($\tau$) space, where we expect the foreground signal to be confined to a region within the ``horizon delay''. This is the maximum delay in the time of arrival of a wavefront between the two antennas of the baseline, which occurs for a source at the horizon. Outside of the horizon delay, we would ideally expect only the 21cm signal and noise to be present, and so can `avoid' foregrounds by analysing only these modes.

To estimate the delay spectra, cross-power spectra are formed by multiplying the delay-transformed visibilities from redundant baselines (same length and orientation), under the idealised assumption that redundant baselines observe the same sky visibilities, plus differences only due to noise.
The cross-power spectra from all pairs of redundant baselines are then averaged at each observing time, then averaged over time, before finally averaging over baselines of the same length (but different orientation). This process follows a similar estimation procedure used in \citet{2022ApJ...925..221A}. In this section we will show the results of this process using unfiltered and filtered visibilities. Regarding Fig.~\ref{pipeline}, we begin at the delay spectrum estimation step.

At this step we apply a delay transform to each baseline at each time, using a suitable tapering function (in this case a Blackman-Harris window), and then  multiply with a suitable weighting defined by the optimal quadratic estimator (OQE) formalism to form the cross-spectra. We follow a similar approach as in recent HERA analyses and use a uniform weighting, which does not produce optimal estimates, but which avoids signal loss issues; see \citet{Pascua} for a more detailed explanation. Note that baselines are not paired with themselves in this analysis, i.e. `auto-baseline pairs' are excluded. 

Following the initial formation of the delay spectra, we average together the cross-spectra 
of redundant baselines, at each time. This is the `incoherent redundant average' step. Following this, the per-redundant-group delay spectra are  averaged in time. Both of these averages can be performed in a way that respect the noise weights of each data sample, but for our simulations the weights are all uniform. 
We do not attempt to model the weighting/covariance between time samples introduced by the fringe-rate filtering procedure here, but note that any filter will induce correlations between previously independent time samples. This is  important for errorbar estimation of the time-averaged delay spectrum. Errorbar estimation for the filtered data is a complex topic; a cursory examination is presented in \citet{2023MNRAS.521.5191W} which needs to be expanded to cover the full HERA array and use realistic fringe-rate parameters. The final step of the pipeline we discuss here incoherently averages delay spectra into bins of the same baseline length.

We will  track the  pair of baselines (3, 6) and (8, 11) through the process, these are both 43.8m East-West baselines in the same redundant group. 

Fig.~\ref{delay_spec_summary_both}  show the delay spectra for our chosen baseline-pair/redundant group for data that have had the mainlobe (left column) and notch (right column) filters applied. For these delay spectra the full delay range is $-5154$ to $5129$ ns, but we show the range $-2000$ to $2000$ ns as only flat noise extends beyond this range. The first row is the raw delay spectrum for the baseline pair without any averaging (apart from the coherent time average that was applied to the visibilities), and at a time half-way through the observation period; the second row is after averaging over the available baseline pairs within the redundant group at the same time; the third row shows the subsequent unweighted  average of the previous step over all times; and the fourth row shows the average of the previous step and all other  redundant/time-averaged cross spectra of the same baseline length. We show both filtered and unfiltered visibilities.
Delay spectra are complex valued and averaged as such, but it is a HERA convention to plot the absolute value of the real component, as is done here, for data and noise. 

For both filters we see that there is some alteration to the power within the horizon, with the peak shifting to 0 ns in the filtered data and reducing the power just outside the peak. The drop in power outside the horizon to the noise level  occurs closer to the horizon for the filtered data, but likely due to other overall lower values for the filtered data. This drop is located well outside the horizon for both filtered/unfiltered cases, indicating a buffer region that should be ignored when examining power beyond the horizon.

After the power drops outside the horizon, the level of noise is fairly flat to the limit of the delay range, indicating we can ignore extreme delays and focus on the wedge for further analysis. The noise level for the filtered delay spectra is always below  that of the unfiltered data for the mainlobe filter, but for the notch they are similar. For the mainlobe filter the noise level is below that of the unfiltered data at the beginning of the process (One baseline), but they tend to the same level towards the end of the process ($|\vec{b}|$ average). This is not seen for the notch filter, where the noise levels are about the same all the way through. Overall, the noise drops from one step to the next as averaging proceeds, but the area in and around the horizon, where foregrounds are present, increases.

\begin{figure*}
\centering
\includegraphics[width=0.66\textwidth]{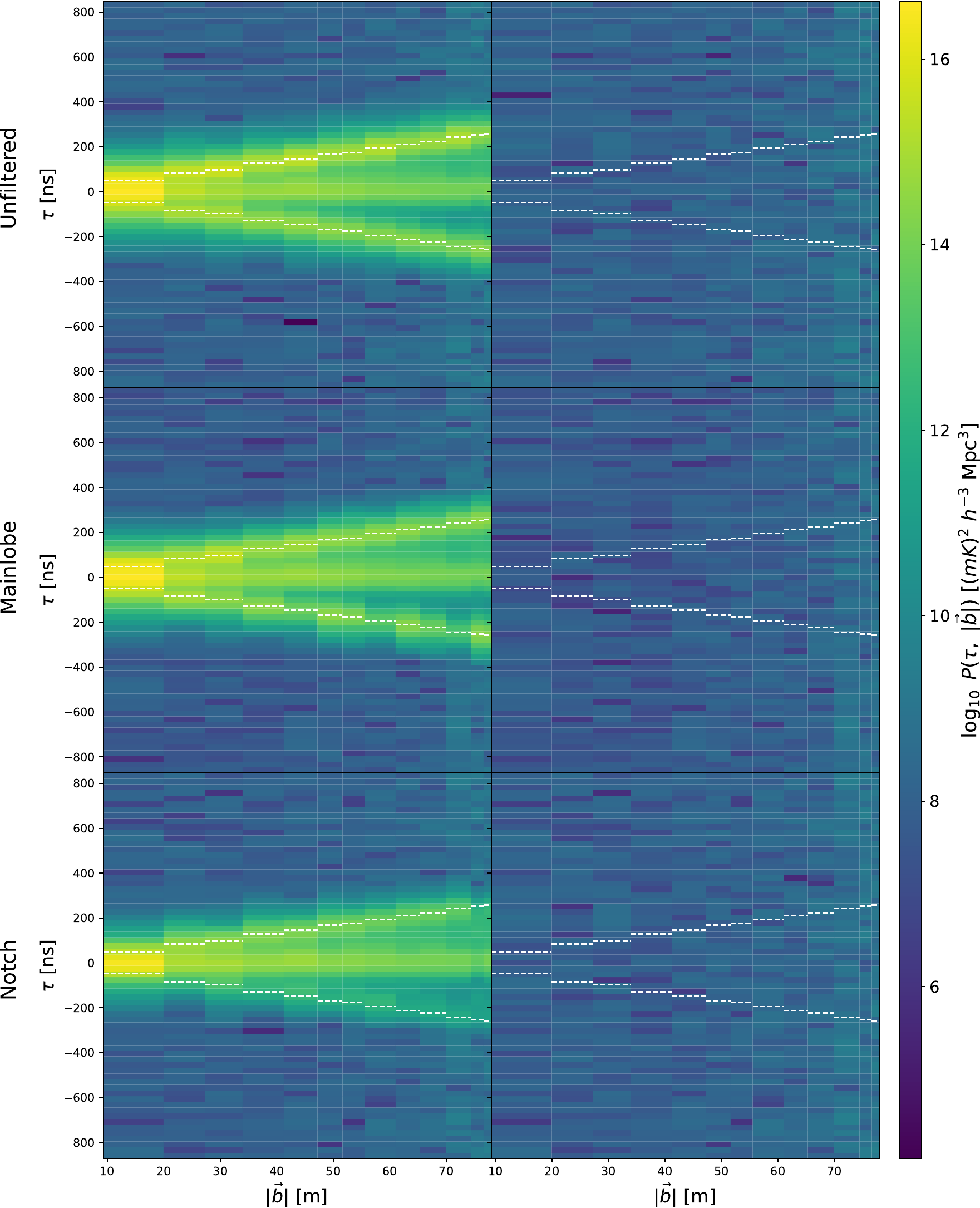}
\caption{{2D power spectra that are the end result of the data processing pipeline. The top row contains the spectra generated from unfiltered visibilities, the middle row contains the spectra generated from mainlobe-filtered visibilities, and the bottom contains the spectra from the notch filter. The left column contains the 2D power spectra, the right column contains the 2D noise power spectra, obtained from passing the noise in the visibilities through the pipeline.}}
\label{wedges}
\end{figure*}

\begin{figure*}
\centering
\includegraphics[width=\textwidth]{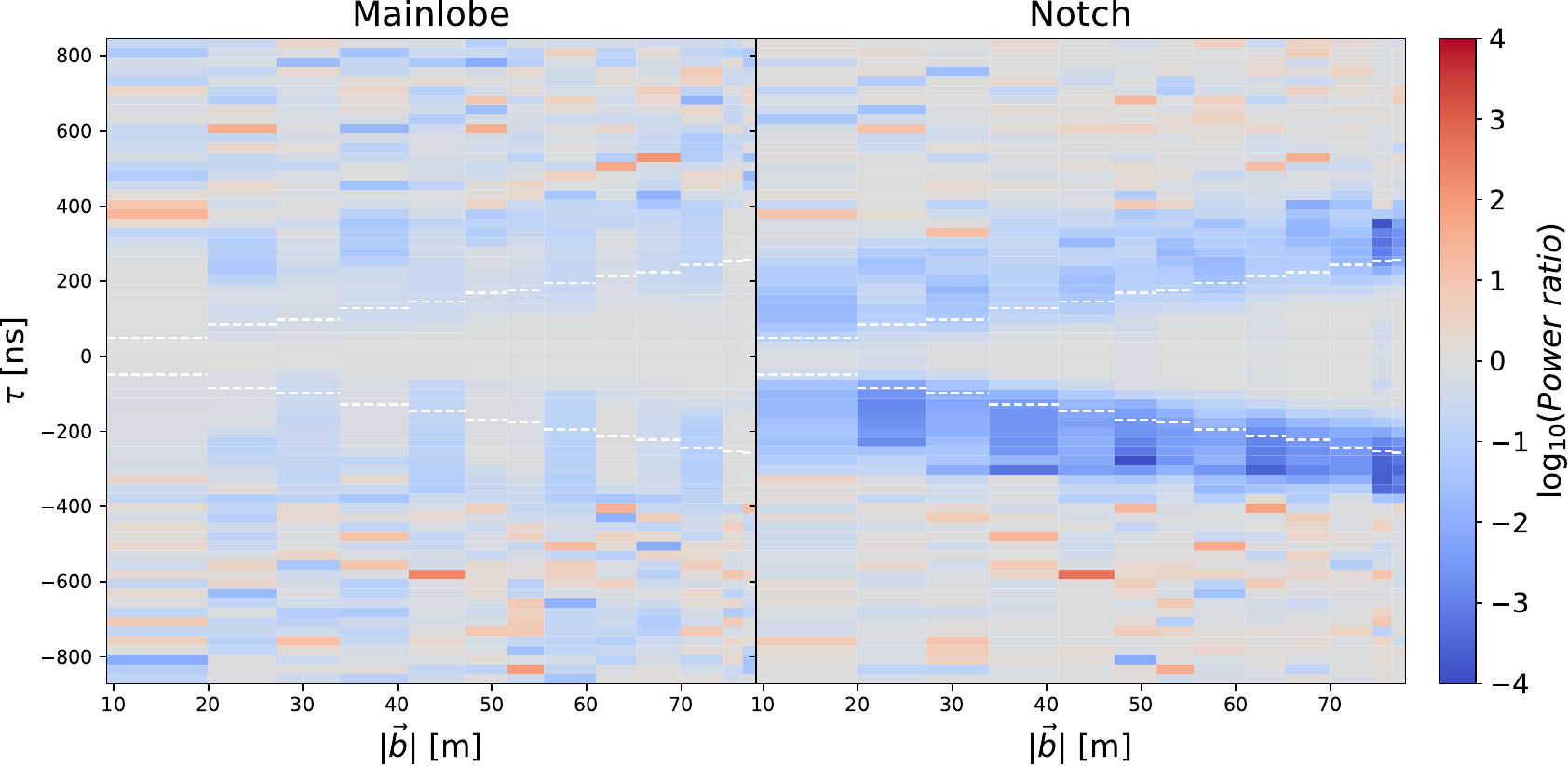}
\caption{Each of these power spectra is the result of dividing  power spectra that are the result of  fringe-rate filtered visibilities, by power spectra produced from unfiltered visibilities. The left spectrum is the mainlobe 2D power spectra (Fig. \ref{wedges} middle left), divided by the unfiltered 2D power spectra (Fig. \ref{wedges} top left). The right spectrum is the same, but for notch filtered visibilities.}
\label{wedges_ratio}
\end{figure*}

\begin{figure*}
\centering
\includegraphics[width=\textwidth]{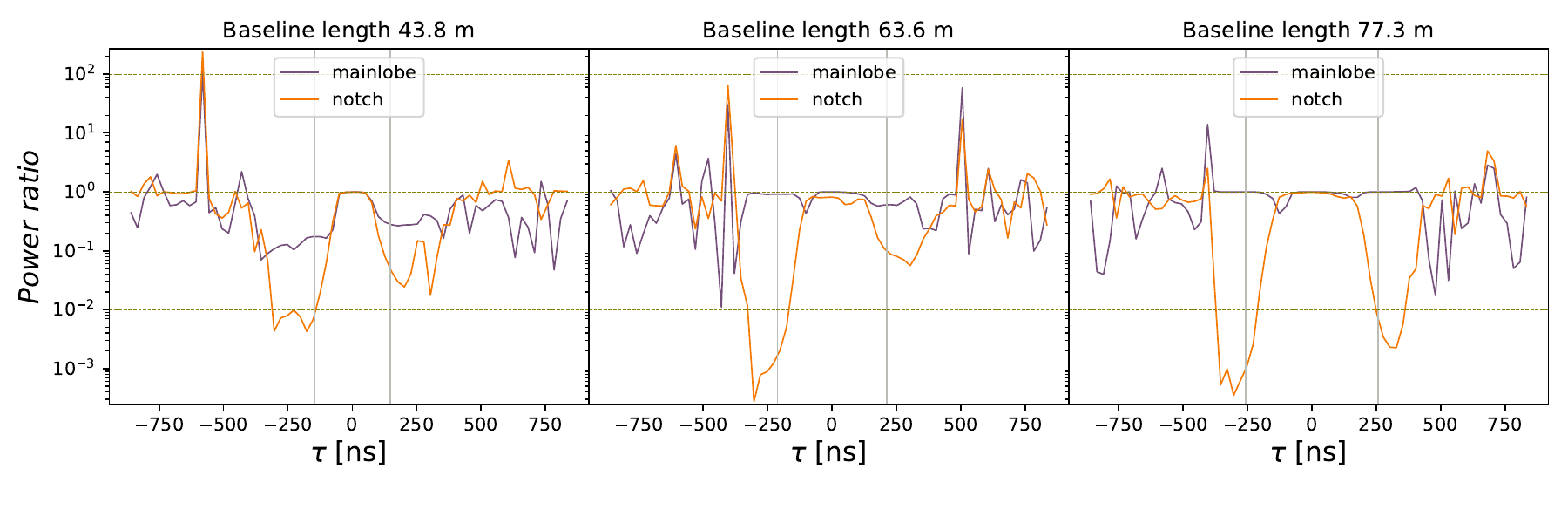}
\caption{Vertical cuts of the 2D power spectrum ratios in Fig. \ref{wedges_ratio}, at different baseline lengths. The vertical grey lines indicate the horizon delay.}
\label{av_wedges_ratio_cuts}
\end{figure*}

\begin{figure*}
\centering
\includegraphics[width=\textwidth]{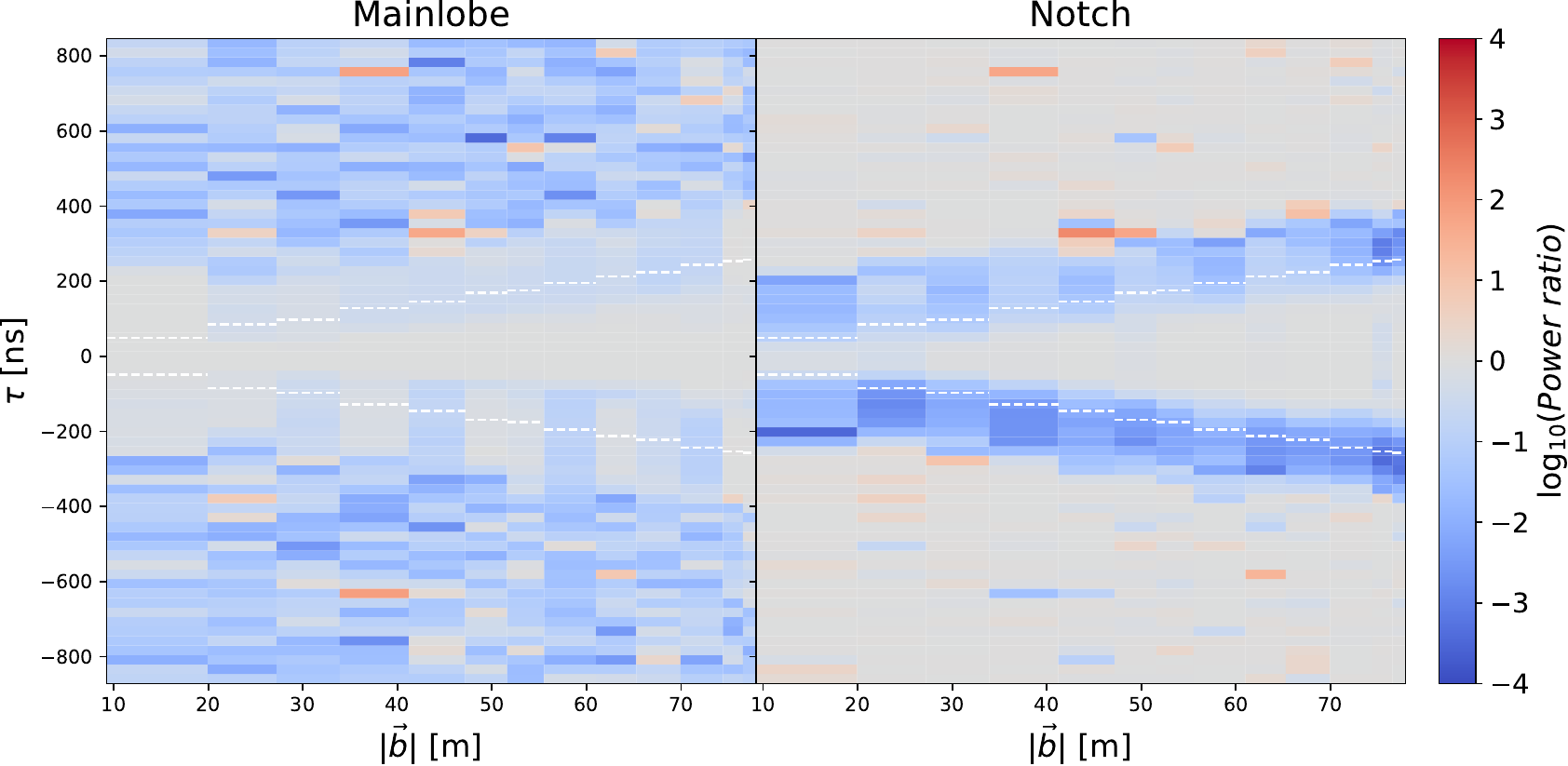}
\caption{As for Fig. \ref{wedges_ratio},  generated from the same simulations and data processing pipeline, except that the coherent averaging step has been skipped. }
\label{wedges_ratio_noav}
\end{figure*}

\begin{figure*}
\centering
\includegraphics[width=\textwidth]{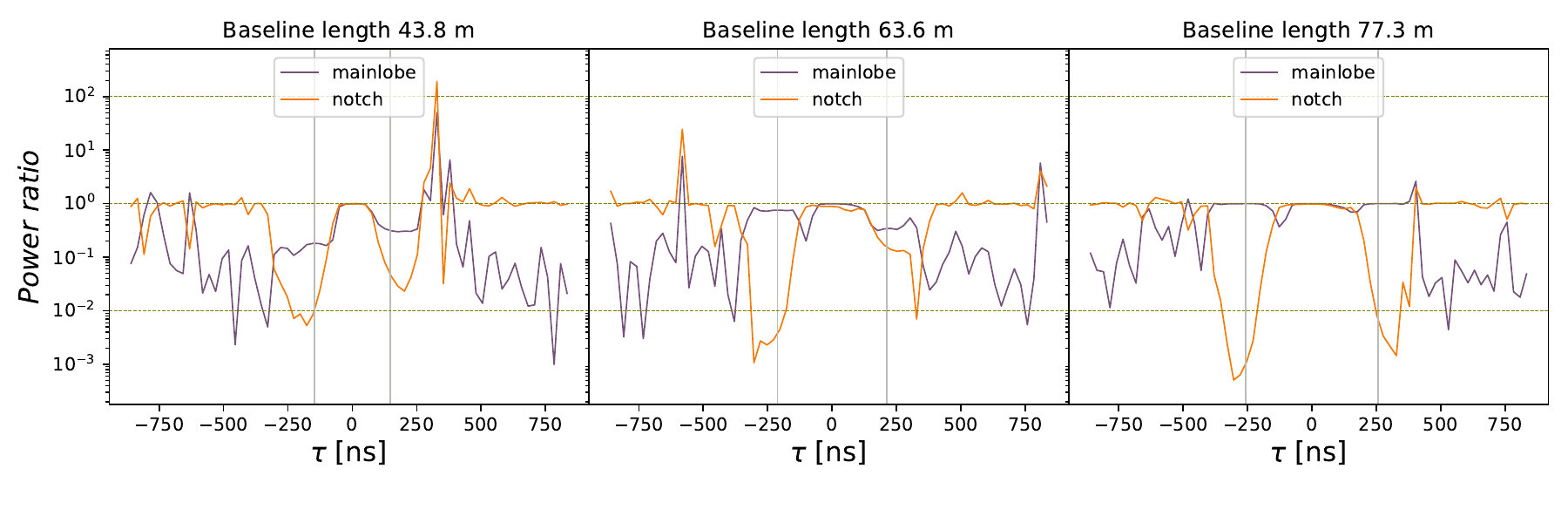}
\caption{As for Fig. \ref{av_wedges_ratio_cuts},  generated from the same simulations and data processing pipeline, except that the coherent averaging step has been skipped. }
\label{wedges_ratio_cuts_noav}
\end{figure*}

\begin{table}
\centering
\begin{tabular}{|c|c|c|}
 & {\bf Visibilities+noise} & {\bf Noise only}  \\
\hline
{\bf Unfiltered}     & $ 1.40\times 10^4 - 4.16\times 10^{16} $  & $1.97\times 10^5 - 3.86\times 10^9$    \\
{\bf Mainlobe }   & $8.50\times 10^5 - 4.13\times 10^{16}$  & $2.26\times 10^5 - 2.10\times 10^9$    \\
{\bf Notch }  & $4.10\times 10^5 - 2.79\times 10^{16}$  & $3.36\times 10^5 - 3.42\times 10^9$ 
\\
\hline
\end{tabular}
\caption{The ranges of power in the 2D power spectra in Fig. \ref{wedges}. Each row and column of numbers corresponds to the power spectra in the same position in Fig. \ref{wedges}.  The first column corresponds to power spectra generated from noise visibilities, and the second column to power spectra generated from noise only, as described in the text. The units are $mK^2\ h^{-3}\  {\rm Mpc}^3$.}
\label{wedge_statistics}
\end{table}

\subsection{Effect of filters on 2D power spectra} \label{sec:res:2dspec}

The delay spectrum from the previous section represents a single baseline length, and there will be several others corresponding to different lengths. Since different baseline lengths approximately measure different Fourier modes of the sky, when they are plotted vertically and stacked horizontally by baseline length, we obtain a 2D power spectrum, the last stage in the pipeline. The horizon delay increases by baseline length, and when plotted on the 2D power spectrum delineates a ``wedge region'' in which the foregrounds are, ideally, contained, and outside of which exists, ideally, the 21cm signal and noise. 

 Fig. \ref{wedges} shows the power spectra generated from the unfiltered, mainlobe-filtered, and notch-filtered visibilities. The first column contains the power spectra generated from noisy visibilities passed through the pipeline, the second column contains power spectra generated from noise only. The noise is extracted from the visibilities and passed through the rest of the pipeline. The horizon delay is indicated by the white lines. Table \ref{wedge_statistics} shows the range of power in each power spectra.

The Figure and Table indicate that the power spectra are quite similar. Fig. \ref{delay_spec_summary_both} (last row) has hinted at the  shape we expect to see in the baseline bins, with a fairly flat noise level and a roll off outside the horizons. 

To more closely examine the difference between the unfiltered and filtered cases we take the ratio of the power spectra,  shown in Fig. \ref{wedges_ratio}. The left ``ratio spectrum'' is the result of dividing the mainlobe 2D power spectra by the unfiltered 2D power spectra. The right ratio spectra is the same, but for the notch filter case. Power ratios below 1 mean that the power is less in the filtered case than the unfiltered. The 2D power ratio spectra are similar although the notch filter shows a significant drop in power just outside the horizon, around fringe-rate 0~mHz, which the mainlobe filter does not. This may seem puzzling, since  the mainlobe filter in our point source example for an EW baseline (Fig. \ref{mainlobe_ew})  would remove fringe-rate 0~MHz. However, the mainlobe filter bounds change depending on baseline orientation, and for  baselines with a significant NS orientation the mainlobe filter can retain fringe-rate 0~mHz.

The ratio spectra also show that there are locations where the ratio is greater than 1, indicating that filtering has produced an increase in power. These locations are all outside the wedge and depend on the random noise added to the visibilities in the simulation;  they may not appear at all, or change location.

To examine the ratios in more detail, Fig. \ref{av_wedges_ratio_cuts} contains vertical cuts of the ratio spectra. Each panel represents a different baseline length. Within a panel, cuts for both the mainlobe and notch cases are indicated by the legend. The wedge horizon for the baseline length is shown by vertical grey lines.  These show that at the horizon there is often a drop in power in the filtered spectra compared to the unfiltered; for example for baseline length 63.6 m there are two sharp drops at the horizon for the notch filter. We expect a drop at the horizon because the fringe-rates are around 0 mHz there, and the notch filter will remove these;  the mainlobe filter will sometimes remove these, depending on the filter bounds, which depends on baseline orientation. We also see some instances where the ratio is above 1, for example for baseline length 63.6 m there are sharp increases just outside both horizons.

The similarity between the power spectra is worthy of more investigation, and it appears that coherent averaging is at least partly the cause. Fig. \ref{wedges_ratio_noav} is generated from the same simulations and data processing pipeline as Fig. \ref{wedges_ratio} except the coherent averaging step has been skipped. Similarly, Fig. \ref{wedges_ratio_cuts_noav} is the same as Fig. \ref{av_wedges_ratio_cuts} with the coherent averaging step  skipped.  The mainlobe case looks similar in both figures (in that the ratios are variable), but the ratio values are at a lower level outside the wedge when coherent averaging is skipped. The notch case has a much more uniform ratio of 1 outside the wedge when coherent averaging is skipped, although there are still instances outside the wedge where the ratio is greater than 1. The uniformity outside the wedge indicates that the unfiltered and filtered power spectra are similar there. It is not surprising that the  different filters have different effects on power spectra, but the interaction between the filters and time averaging requires further investigation.
\begin{figure}
\centering
\includegraphics[width=\columnwidth]{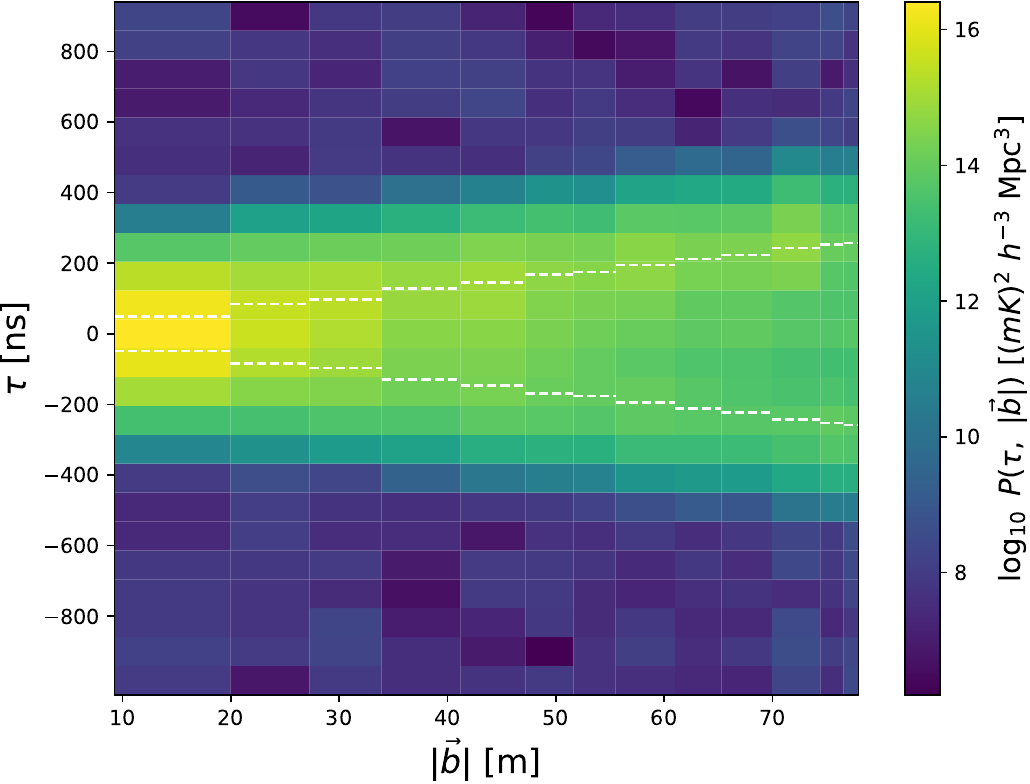}
\caption{Shows the unfiltered power spectra that is produced from a reduced frequency band of 109.79--121.92 ~MHz. It is visually similar to the power spectra in Fig. \ref{wedges}, top left, but with the wedge widened due to the change in band.}
\label{band_wedges}
\end{figure}

\begin{figure*}
\centering
\includegraphics[width=\textwidth]{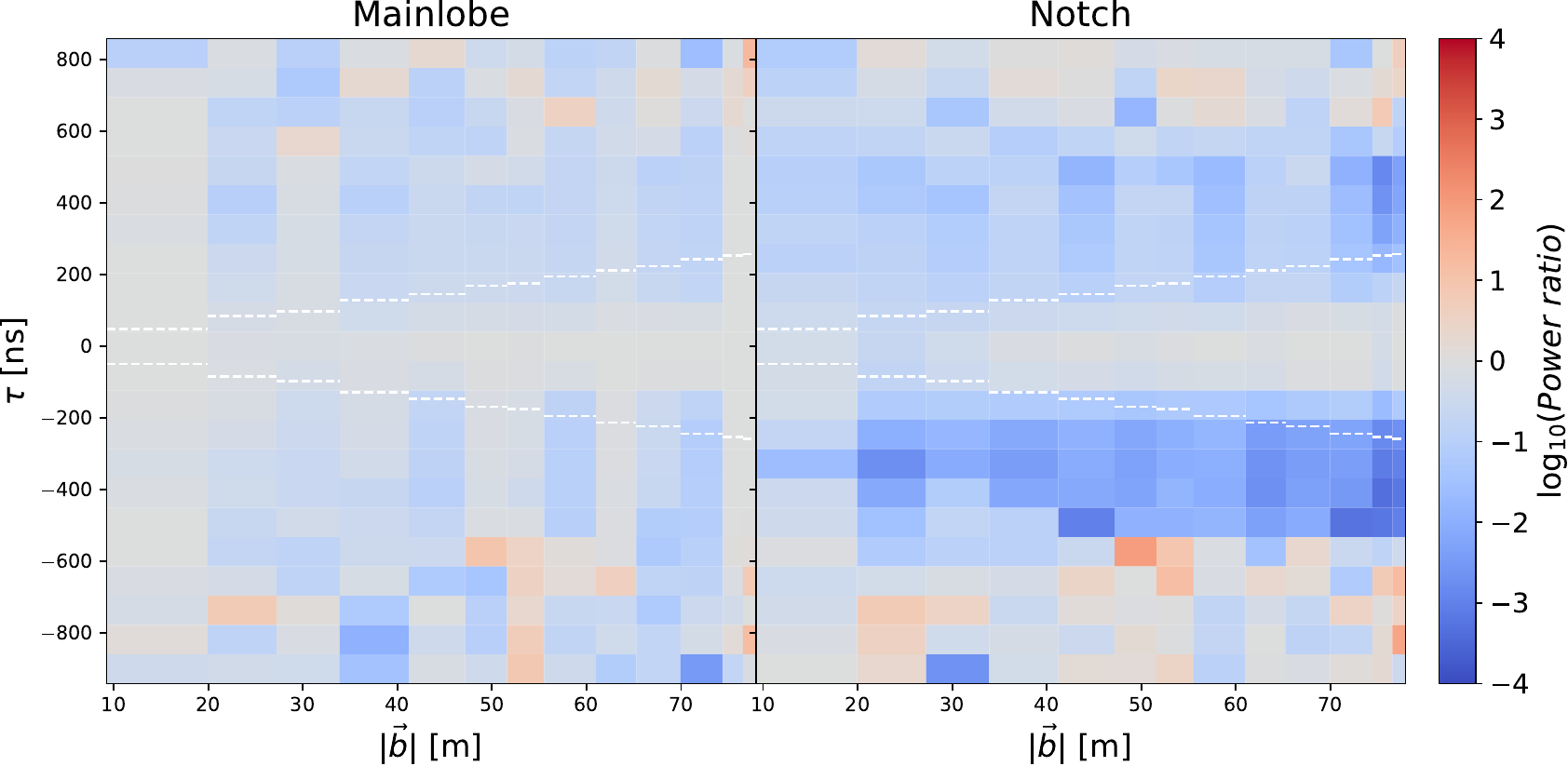}
\caption{As for Fig. \ref{wedges_ratio} but using the 2D power spectra from a reduced frequency band.}
\label{band_wedges_ratio}
\end{figure*}

\begin{figure*}
\centering
\includegraphics[width=\textwidth]{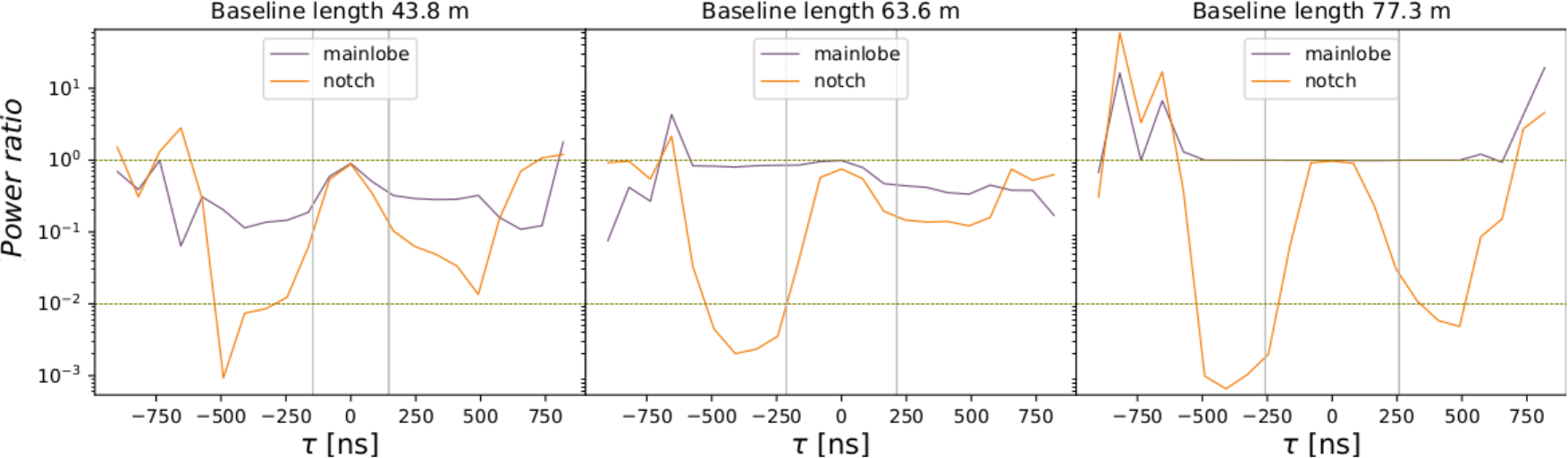}
\caption{As for Fig. \ref{av_wedges_ratio_cuts} but the cuts are taken from the power spectra ratio plots in Fig. \ref{band_wedges_ratio}, which used a reduced frequency band.}
\label{band_av_wedges_ratio_cuts}
\end{figure*}

\subsection{Using more realistic HERA-like observations}

The above results and analysis are based on simulations that use a frequency range of 100--140 MHz,  $ z \approx 9.1-13.2$, during which time the  Universe will be evolving. To capture power spectra at a particular instant during the Universe's evolution, we should use a smaller redshift range, while at the same time providing enough redshift bins for resolution of the spectra. When processing real observations, HERA will take power spectra from several frequency bands of width approximately 10~MHz,  in this section we will use only the frequency range of HERA Band 4, 109.79--121.92~MHz, $ z \approx 10.6-11.9$. 

We also change the noise to simulate LST binning of visibilities over 100 days; LST binning will be used in the HERA pipeline and was part of the pipeline from which we extracted Fig. \ref{pipeline}. At a particular  LST we assume that the noiseless visibility will be the same every day, but random noise will be different. If the noisy visibility is averaged over those days, the noiseless visibility remains the same, but the variance of the noise, on average, reduces by a factor of 100. Therefore, we reduce the variance of the noise added to the visibilities by  a factor of 100.

The power spectra that are produced from the reduced frequency band look almost the same as those in Fig. \ref{wedges} except that the wedge is wider (due to the different band). For that reason we only show the unfiltered power spectra of the reduced band as an example, in Fig.~\ref{band_wedges}. This shows that there is considerable leakage outside the wedge when the narrower band is used. Of more interest are the ratio plots (Fig.~\ref{band_wedges_ratio}), and vertical cuts through the ratio plots (Fig.~\ref{band_av_wedges_ratio_cuts}). The ratios of the filtered/unfiltered power spectra, Fig.~\ref{band_wedges_ratio}, still show that the notch filter produces a significant dip in power outside the wedge that the mainlobe filter does not. Fig.~\ref{band_av_wedges_ratio_cuts}, containing vertical cuts through the ratio power spectra, shows the dips produced by the notch filter more clearly.

\section{Conclusions} \label{sec:conclusions}

The Hydrogen Epoch of Reionization Array (HERA) is now the most sensitive radio interferometer targeting the 21cm brightness temperature fluctuation signal at high redshift \citep{2023ApJ...945..124H}, and will likely remain so until the SKAO-LOW instrument comes online in several years from now. Forthcoming seasons of HERA data will unavoidably require Fourier filtering in both the frequency (delay) and time (fringe-rate) dimensions, due to the presence of foreground contamination and a variety of systematic effects such as mutual coupling that are relatively localised in these domains. The fringe-rate filters in particular can seem unintuitive, as they remove power in a way that is spread across the time axis, causing correlations between time samples that weren't previously there, as well as introducing artifacts such as ringing and potentially some degree of signal loss. This interferes with traditional data analysis approaches that tend to split observations into smaller time windows (`fields') and average over time samples to build up sensitivity. The loss of intuition for how these averaging operations should work -- that we are no longer able to simply combine independent data samples -- risks making the analyses harder to interpret, or even leading to analysis choices that introduce errors, for instance in estimating the error bars or determining robust upper limits.

In this paper, we have used simple examples and a simplified mock data analysis pipeline applied to semi-realistic visibility simulations to build intuition into what the fringe-rate filters are actually doing to the data at each step of the pipeline. We consider two types of filter, both of them tophat-like -- a mainlobe filter that retains parts of the signal with fringe-rates that fall only within the mainlobe of the primary beam, and a notch filter that removes fringe rates around $f \approx 0$~mHz corresponding to signals that are locked to the ground rather than rotating with the sky, which is symptomatic of certain classes of systematic effects.

We demonstrated the use of the filters on a simple setup of an East-West baseline on the equator, showing that the filters do as intended. The mainlobe filter alters fringe-rates to remove sky locations outside the primary beam,  However, Fig. \ref{mainlobe_ew} shows that the profile of the resulting visibilities over time only approximately matches the beam profile in Fig. \ref{fig:layout} (bottom), which it ideally should. Whether this is due to fringe-rate resolution, the manner of calculating the tophat width and location, the use of a tophat filter, the use of discrete  prolate spheroidal sequences, or other factors, requires investigation, if the ideal is to be reached. It may be that the simple situation of a point source generating slowing changing perfect fringes with no amplitude deviations, is unsuitable for the intended application. In a real-world scenario, the fringe-rates of a baseline will have a more complex structure, generated from variable intensity sources all over the sky. 

The fringe-rate filters applied to a real-world simulation demonstrate this complex structure (e.g. Fig. \ref{mainlobe_vis_summary}). The visibilities in this case contain realistic noise, which is smoothed over time and reduced by the filter. In this case we have multiple frequencies which allow us to make delay transforms and observe the effect of the filter in delay space for a single time. While the fringe-rate filters operate over time, the visibilities are changed over frequency,  altering the power within the horizon delays, adding power to some delays and reducing others. Of interest is that the peak of the delays within the horizon is shifted to $\tau = 0$ ns if it is off 0 ns in the unfiltered data.

The use of discrete prolate spheroidal sequences shows that they do not function as a true tophat filter with a sharp edge, instead leaking outside the bounds of the tophat. Research is suggesting methods to deal with this, such as lessening the width of the tophat so that the leakage is within the filter width originally intended \citep{Pascua}. There are also parameters that can be applied to the use of these sequences, which we have not experimented with, that may mitigate the leakage.

The delay spectra through various averaging stages are surprisingly similar, indicating a peak at 0 ns delay, a drop-off outside the horizon, and a fairly flat noise floor beyond the horizon. What does change is the level of the noise floor, being reduced by the use of fringe-rate filtering. The noise, as expected, also reduces as more delay spectra are averaged at the various stages.

The 2D power spectra are also similar for both filters. To compare 2D power spectra from unfiltered and filtered visibilities, we ratio them, showing that the power outside the wedge drops more compared to the power inside the wedge, when filtering is used. The ratio also shows that outside the wedge, the power can sometimes increase compared to the filtered data, this is not everywhere, but at isolated points within the power spectrum.

In looking for the effects of the filters, we found that there can be a sharp drop in power at the horizon, more than anywhere else in delay space. This is an effect that we expect, particularly from the notch filter, which removes fringe-rates corresponding to emission from the horizon.

The similarities between the power spectra generated by the filters may be partly (or wholly) due to coherent averaging of visibilities after fringe-rate filtering. If coherent averaging is not used, then, at least comparing ratios of power spectra, the power outside the wedge  differs depending on the filter. It therefore shows that the interaction of the filters with coherent averaging, and indeed any subsequent processing steps, should studied further.

In all the data products analysed, from visibilities to power spectra, 
it is clear that there is significant signal loss, simply due to removal of fringe-rates. Signal loss is not the subject of this paper and will be reported separately \citep{Pascua}; developing analytic methods to generate expected signal loss that can be combined with the analysis of errors on 2D power spectra \citep{2021ApJS..255...26T}, for example.
\section*{Acknowledgements}

This result is part of a project that has received funding from the European Research Council (ERC) under the European Union's Horizon 2020 research and innovation programme (Grant agreement No. 948764; PB and MJW). HG and PB acknowledge support from STFC Grant ST/T000341/1.

\section*{Data Availability}
 
The total size of data generated for this paper is approximately 24~TB, if one includes all intermediate data products and the full output of the pipeline for the all the alternative versions -- with coherent averaging or without coherent averaging, different eigenvalue cutoffs, and different frequency ranges. While it is not practical to release these data in full online, we can can supply components of it on request. The software used to generate the simulated data is public, and the configuration files and analysis scripts are also available on request.

\balance


\bibliographystyle{mnras}
\bibliography{frfcomparison}

\begin{thebibliography}{}
\makeatletter
\relax
\def\mn@urlcharsother{\let\do\@makeother \do\$\do\&\do\#\do\^\do\_\do\%\do\~}
\def\mn@doi{\begingroup\mn@urlcharsother \@ifnextchar [ {\mn@doi@}
  {\mn@doi@[]}}
\def\mn@doi@[#1]#2{\def\@tempa{#1}\ifx\@tempa\@empty \href
  {http://dx.doi.org/#2} {doi:#2}\else \href {http://dx.doi.org/#2} {#1}\fi
  \endgroup}
\def\mn@eprint#1#2{\mn@eprint@#1:#2::\@nil}
\def\mn@eprint@arXiv#1{\href {http://arxiv.org/abs/#1} {{\tt arXiv:#1}}}
\def\mn@eprint@dblp#1{\href {http://dblp.uni-trier.de/rec/bibtex/#1.xml}
  {dblp:#1}}
\def\mn@eprint@#1:#2:#3:#4\@nil{\def\@tempa {#1}\def\@tempb {#2}\def\@tempc
  {#3}\ifx \@tempc \@empty \let \@tempc \@tempb \let \@tempb \@tempa \fi \ifx
  \@tempb \@empty \def\@tempb {arXiv}\fi \@ifundefined
  {mn@eprint@\@tempb}{\@tempb:\@tempc}{\expandafter \expandafter \csname
  mn@eprint@\@tempb\endcsname \expandafter{\@tempc}}}

\bibitem[\protect\citeauthoryear{{Aguirre} et~al.,}{{Aguirre}
  et~al.}{2022}]{2022ApJ...924...85A}
{Aguirre} J.~E.,  et~al., 2022, \mn@doi [\apj] {10.3847/1538-4357/ac32cd},
  \href {https://ui.adsabs.harvard.edu/abs/2022ApJ...924...85A} {924, 85}

\bibitem[\protect\citeauthoryear{{Ali} et~al.,}{{Ali}
  et~al.}{2015}]{2015ApJ...809...61A}
{Ali} Z.~S.,  et~al., 2015, \mn@doi [\apj] {10.1088/0004-637X/809/1/61}, \href
  {https://ui.adsabs.harvard.edu/abs/2015ApJ...809...61A} {809, 61}

\bibitem[\protect\citeauthoryear{{Alonso}, {Bull}, {Ferreira}  \&
  {Santos}}{{Alonso} et~al.}{2015}]{2015MNRAS.447..400A}
{Alonso} D.,  {Bull} P.,  {Ferreira} P.~G.,   {Santos} M.~G.,  2015, \mn@doi
  [\mnras] {10.1093/mnras/stu2474}, \href
  {https://ui.adsabs.harvard.edu/abs/2015MNRAS.447..400A} {447, 400}

\bibitem[\protect\citeauthoryear{{Barry}, {Hazelton}, {Sullivan}, {Morales}  \&
  {Pober}}{{Barry} et~al.}{2016}]{2016MNRAS.461.3135B}
{Barry} N.,  {Hazelton} B.,  {Sullivan} I.,  {Morales} M.~F.,   {Pober} J.~C.,
  2016, \mn@doi [\mnras] {10.1093/mnras/stw1380}, \href
  {https://ui.adsabs.harvard.edu/abs/2016MNRAS.461.3135B} {461, 3135}

\bibitem[\protect\citeauthoryear{{Charles}, {Kern}, {Bernardi}, {Bester},
  {Smirnov}, {Fagnoni}  \& {Acedo}}{{Charles}
  et~al.}{2023}]{2023MNRAS.522.1009C}
{Charles} N.,  {Kern} N.,  {Bernardi} G.,  {Bester} L.,  {Smirnov} O.,
  {Fagnoni} N.,   {Acedo} E. d.~L.,  2023, \mn@doi [\mnras]
  {10.1093/mnras/stad1046}, \href
  {https://ui.adsabs.harvard.edu/abs/2023MNRAS.522.1009C} {522, 1009}

\bibitem[\protect\citeauthoryear{{Charles} et~al.,}{{Charles}
  et~al.}{2024}]{2024arXiv240720923C}
{Charles} N.,  et~al., 2024, \mn@doi [arXiv e-prints]
  {10.48550/arXiv.2407.20923}, \href
  {https://ui.adsabs.harvard.edu/abs/2024arXiv240720923C} {p. arXiv:2407.20923}

\bibitem[\protect\citeauthoryear{{Chen} \& {Pullen}}{{Chen} \&
  {Pullen}}{2022}]{2022MNRAS.512.4262C}
{Chen} C.,  {Pullen} A.~R.,  2022, \mn@doi [\mnras] {10.1093/mnras/stac743},
  \href {https://ui.adsabs.harvard.edu/abs/2022MNRAS.512.4262C} {512, 4262}

\bibitem[\protect\citeauthoryear{Cheng, Chang, Bock, Bradford  \& Cooray}{Cheng
  et~al.}{2016}]{Cheng:2016yvu}
Cheng Y.-T.,  Chang T.-C.,  Bock J.,  Bradford C.~M.,   Cooray A.,  2016,
  \mn@doi [Astrophys. J.] {10.3847/0004-637X/832/2/165}, 832, 165

\bibitem[\protect\citeauthoryear{{Cheng} et~al.,}{{Cheng}
  et~al.}{2018}]{2018ApJ...868...26C}
{Cheng} C.,  et~al., 2018, \mn@doi [\apj] {10.3847/1538-4357/aae833}, \href
  {https://ui.adsabs.harvard.edu/abs/2018ApJ...868...26C} {868, 26}

\bibitem[\protect\citeauthoryear{{Choudhuri}, {Bull}  \& {Garsden}}{{Choudhuri}
  et~al.}{2021}]{2021MNRAS.506.2066C}
{Choudhuri} S.,  {Bull} P.,   {Garsden} H.,  2021, \mn@doi [\mnras]
  {10.1093/mnras/stab1795}, \href
  {https://ui.adsabs.harvard.edu/abs/2021MNRAS.506.2066C} {506, 2066}

\bibitem[\protect\citeauthoryear{{DeBoer} et~al.,}{{DeBoer}
  et~al.}{2017}]{2017PASP..129d5001D}
{DeBoer} D.~R.,  et~al., 2017, \mn@doi [\pasp]
  {10.1088/1538-3873/129/974/045001}, \href
  {https://ui.adsabs.harvard.edu/abs/2017PASP..129d5001D} {129, 045001}

\bibitem[\protect\citeauthoryear{{Dor{\'e}} et~al.,}{{Dor{\'e}}
  et~al.}{2016}]{2016arXiv160607039D}
{Dor{\'e}} O.,  et~al., 2016, \mn@doi [arXiv e-prints]
  {10.48550/arXiv.1606.07039}, \href
  {https://ui.adsabs.harvard.edu/abs/2016arXiv160607039D} {p. arXiv:1606.07039}

\bibitem[\protect\citeauthoryear{{Ewall-Wice}, {Dillon}, {Liu}  \&
  {Hewitt}}{{Ewall-Wice} et~al.}{2017}]{2017MNRAS.470.1849E}
{Ewall-Wice} A.,  {Dillon} J.~S.,  {Liu} A.,   {Hewitt} J.,  2017, \mn@doi
  [\mnras] {10.1093/mnras/stx1221}, \href
  {https://ui.adsabs.harvard.edu/abs/2017MNRAS.470.1849E} {470, 1849}

\bibitem[\protect\citeauthoryear{{Ewall-Wice} et~al.,}{{Ewall-Wice}
  et~al.}{2021}]{2021MNRAS.500.5195E}
{Ewall-Wice} A.,  et~al., 2021, \mn@doi [\mnras] {10.1093/mnras/staa3293},
  \href {https://ui.adsabs.harvard.edu/abs/2021MNRAS.500.5195E} {500, 5195}

\bibitem[\protect\citeauthoryear{{Fagnoni}, {de Lera Acedo}, {Drought},
  {DeBoer}, {Riley}, {Razavi-Ghods}, {Carey}  \& {Parsons}}{{Fagnoni}
  et~al.}{2021}]{2021ITAP...69.8143F}
{Fagnoni} N.,  {de Lera Acedo} E.,  {Drought} N.,  {DeBoer} D.~R.,  {Riley} D.,
   {Razavi-Ghods} N.,  {Carey} S.,   {Parsons} A.~R.,  2021, \mn@doi [IEEE
  Transactions on Antennas and Propagation] {10.1109/TAP.2021.3083788}, \href
  {https://ui.adsabs.harvard.edu/abs/2021ITAP...69.8143F} {69, 8143}

\bibitem[\protect\citeauthoryear{{Fudamoto} et~al.,}{{Fudamoto}
  et~al.}{2021}]{2021Natur.597..489F}
{Fudamoto} Y.,  et~al., 2021, \mn@doi [\nat] {10.1038/s41586-021-03846-z},
  \href {https://ui.adsabs.harvard.edu/abs/2021Natur.597..489F} {597, 489}

\bibitem[\protect\citeauthoryear{{Furlanetto}, {Oh}  \& {Briggs}}{{Furlanetto}
  et~al.}{2006}]{furlanetto06}
{Furlanetto} S.~R.,  {Oh} S.~P.,   {Briggs} F.~H.,  2006, \mn@doi [\physrep]
  {10.1016/j.physrep.2006.08.002}, \href
  {https://ui.adsabs.harvard.edu/abs/2006PhR...433..181F} {433, 181}

\bibitem[\protect\citeauthoryear{{Gong}, {Chen}  \& {Cooray}}{{Gong}
  et~al.}{2020}]{2020ApJ...894..152G}
{Gong} Y.,  {Chen} X.,   {Cooray} A.,  2020, \mn@doi [\apj]
  {10.3847/1538-4357/ab87a0}, \href
  {https://ui.adsabs.harvard.edu/abs/2020ApJ...894..152G} {894, 152}

\bibitem[\protect\citeauthoryear{Gruenbacher \& Hummels}{Gruenbacher \&
  Hummels}{1994}]{330397}
Gruenbacher D.,  Hummels D.,  1994, \mn@doi [IEEE Transactions on Signal
  Processing] {10.1109/78.330397}, 42, 3276

\bibitem[\protect\citeauthoryear{{HERA Collaboration}}{{HERA
  Collaboration}}{2022}]{2022ApJ...925..221A}
{HERA Collaboration} 2022, \mn@doi [\apj] {10.3847/1538-4357/ac1c78}, \href
  {https://ui.adsabs.harvard.edu/abs/2022ApJ...925..221A} {925, 221}

\bibitem[\protect\citeauthoryear{{HERA Collaboration} et~al.,}{{HERA
  Collaboration} et~al.}{2023}]{2023ApJ...945..124H}
{HERA Collaboration} et~al., 2023, \mn@doi [\apj] {10.3847/1538-4357/acaf50},
  \href {https://ui.adsabs.harvard.edu/abs/2023ApJ...945..124H} {945, 124}

\bibitem[\protect\citeauthoryear{{Harris}}{{Harris}}{1978}]{1978IEEEP..66...51H}
{Harris} F.~J.,  1978, IEEE Proceedings, \href
  {https://ui.adsabs.harvard.edu/abs/1978IEEEP..66...51H} {66, 51}

\bibitem[\protect\citeauthoryear{{Helmboldt}, {Kooi}, {Ray}, {Clarke},
  {Intema}, {Kassim}  \& {Mroczkowski}}{{Helmboldt}
  et~al.}{2019}]{2019RaSc...54.1002H}
{Helmboldt} J.~F.,  {Kooi} J.~E.,  {Ray} P.~S.,  {Clarke} T.~E.,  {Intema}
  H.~T.,  {Kassim} N.~E.,   {Mroczkowski} T.,  2019, \mn@doi [Radio Science]
  {10.1029/2019RS006887}, \href
  {https://ui.adsabs.harvard.edu/abs/2019RaSc...54.1002H} {54, 1002}

\bibitem[\protect\citeauthoryear{{Hurley-Walker} et~al.,}{{Hurley-Walker}
  et~al.}{2017}]{2017MNRAS.464.1146H}
{Hurley-Walker} N.,  et~al., 2017, \mn@doi [\mnras] {10.1093/mnras/stw2337},
  \href {https://ui.adsabs.harvard.edu/abs/2017MNRAS.464.1146H} {464, 1146}

\bibitem[\protect\citeauthoryear{{Jensen}, {Majumdar}, {Mellema}, {Lidz},
  {Iliev}  \& {Dixon}}{{Jensen} et~al.}{2016}]{2016MNRAS.456...66J}
{Jensen} H.,  {Majumdar} S.,  {Mellema} G.,  {Lidz} A.,  {Iliev} I.~T.,
  {Dixon} K.~L.,  2016, \mn@doi [\mnras] {10.1093/mnras/stv2679}, \href
  {https://ui.adsabs.harvard.edu/abs/2016MNRAS.456...66J} {456, 66}

\bibitem[\protect\citeauthoryear{{Kennedy}, {Bull}, {Wilensky}, {Burba}  \&
  {Choudhuri}}{{Kennedy} et~al.}{2023}]{2023ApJS..266...23K}
{Kennedy} F.,  {Bull} P.,  {Wilensky} M.~J.,  {Burba} J.,   {Choudhuri} S.,
  2023, \mn@doi [\apjs] {10.3847/1538-4365/acc324}, \href
  {https://ui.adsabs.harvard.edu/abs/2023ApJS..266...23K} {266, 23}

\bibitem[\protect\citeauthoryear{{Kern}, {Parsons}, {Dillon}, {Lanman},
  {Fagnoni}  \& {de Lera Acedo}}{{Kern} et~al.}{2019}]{2019ApJ...884..105K}
{Kern} N.~S.,  {Parsons} A.~R.,  {Dillon} J.~S.,  {Lanman} A.~E.,  {Fagnoni}
  N.,   {de Lera Acedo} E.,  2019, \mn@doi [\apj] {10.3847/1538-4357/ab3e73},
  \href {https://ui.adsabs.harvard.edu/abs/2019ApJ...884..105K} {884, 105}

\bibitem[\protect\citeauthoryear{{Kern} et~al.,}{{Kern} et~al.}{2020}]{Kern+20}
{Kern} N.~S.,  et~al., 2020, \mn@doi [\apj] {10.3847/1538-4357/ab5e8a}, \href
  {https://ui.adsabs.harvard.edu/abs/2020ApJ...888...70K} {888, 70}

\bibitem[\protect\citeauthoryear{{Kittiwisit} et~al.,}{{Kittiwisit}
  et~al.}{2023}]{2023arXiv231209763K}
{Kittiwisit} P.,  et~al., 2023, \mn@doi [arXiv e-prints]
  {10.48550/arXiv.2312.09763}, \href
  {https://ui.adsabs.harvard.edu/abs/2023arXiv231209763K} {p. arXiv:2312.09763}

\bibitem[\protect\citeauthoryear{{Kolopanis} et~al.,}{{Kolopanis}
  et~al.}{2019}]{2019ApJ...883..133K}
{Kolopanis} M.,  et~al., 2019, \mn@doi [\apj] {10.3847/1538-4357/ab3e3a}, \href
  {https://ui.adsabs.harvard.edu/abs/2019ApJ...883..133K} {883, 133}

\bibitem[\protect\citeauthoryear{{Kovetz} et~al.,}{{Kovetz}
  et~al.}{2017}]{2017arXiv170909066K}
{Kovetz} E.~D.,  et~al., 2017, \mn@doi [arXiv e-prints]
  {10.48550/arXiv.1709.09066}, \href
  {https://ui.adsabs.harvard.edu/abs/2017arXiv170909066K} {p. arXiv:1709.09066}

\bibitem[\protect\citeauthoryear{{Lidz} \& {Taylor}}{{Lidz} \&
  {Taylor}}{2016}]{2016ApJ...825..143L}
{Lidz} A.,  {Taylor} J.,  2016, \mn@doi [\apj] {10.3847/0004-637X/825/2/143},
  \href {https://ui.adsabs.harvard.edu/abs/2016ApJ...825..143L} {825, 143}

\bibitem[\protect\citeauthoryear{{Liu}, {Tegmark}, {Bowman}, {Hewitt}  \&
  {Zaldarriaga}}{{Liu} et~al.}{2009}]{2009MNRAS.398..401L}
{Liu} A.,  {Tegmark} M.,  {Bowman} J.,  {Hewitt} J.,   {Zaldarriaga} M.,  2009,
  \mn@doi [\mnras] {10.1111/j.1365-2966.2009.15156.x}, \href
  {https://ui.adsabs.harvard.edu/abs/2009MNRAS.398..401L} {398, 401}

\bibitem[\protect\citeauthoryear{{Liu}, {Parsons}  \& {Trott}}{{Liu}
  et~al.}{2014}]{2014PhRvD..90b3018L}
{Liu} A.,  {Parsons} A.~R.,   {Trott} C.~M.,  2014, \mn@doi [\prd]
  {10.1103/PhysRevD.90.023018}, \href
  {https://ui.adsabs.harvard.edu/abs/2014PhRvD..90b3018L} {90, 023018}

\bibitem[\protect\citeauthoryear{{Liu}, {Newburgh}, {Saliwanchik}  \&
  {Slosar}}{{Liu} et~al.}{2022}]{2022arXiv220307864L}
{Liu} A.,  {Newburgh} L.,  {Saliwanchik} B.,   {Slosar} A.,  2022, \mn@doi
  [arXiv e-prints] {10.48550/arXiv.2203.07864}, \href
  {https://ui.adsabs.harvard.edu/abs/2022arXiv220307864L} {p. arXiv:2203.07864}

\bibitem[\protect\citeauthoryear{Mathews, Breakall  \& Karawas}{Mathews
  et~al.}{1985}]{mathews}
Mathews J.,  Breakall J.,   Karawas G.,  1985, \mn@doi [IEEE Transactions on
  Acoustics, Speech, and Signal Processing] {10.1109/TASSP.1985.1164732}, 33,
  1471

\bibitem[\protect\citeauthoryear{{Mellema} et~al.,}{{Mellema}
  et~al.}{2013}]{mellema13}
{Mellema} G.,  et~al., 2013, \mn@doi [Experimental Astronomy]
  {10.1007/s10686-013-9334-5}, \href
  {https://ui.adsabs.harvard.edu/abs/2013ExA....36..235M} {36, 235}

\bibitem[\protect\citeauthoryear{{Mirocha} et~al.,}{{Mirocha}
  et~al.}{2019}]{2019arXiv190306218M}
{Mirocha} J.,  et~al., 2019, \mn@doi [arXiv e-prints]
  {10.48550/arXiv.1903.06218}, \href
  {https://ui.adsabs.harvard.edu/abs/2019arXiv190306218M} {p. arXiv:1903.06218}

\bibitem[\protect\citeauthoryear{{Moore} \& {Cada}}{{Moore} \&
  {Cada}}{2004}]{moore}
{Moore} I.~C.,  {Cada} M.,  2004, Appl. Comput. Harmon. Anal., 16, 208

\bibitem[\protect\citeauthoryear{{Morales} \& {Wyithe}}{{Morales} \&
  {Wyithe}}{2010}]{morales10}
{Morales} M.~F.,  {Wyithe} J. S.~B.,  2010, \mn@doi [\araa]
  {10.1146/annurev-astro-081309-130936}, \href
  {https://ui.adsabs.harvard.edu/abs/2010ARA&A..48..127M} {48, 127}

\bibitem[\protect\citeauthoryear{{Morales}, {Hazelton}, {Sullivan}  \&
  {Beardsley}}{{Morales} et~al.}{2012}]{2012ApJ...752..137M}
{Morales} M.~F.,  {Hazelton} B.,  {Sullivan} I.,   {Beardsley} A.,  2012,
  \mn@doi [\apj] {10.1088/0004-637X/752/2/137}, \href
  {https://ui.adsabs.harvard.edu/abs/2012ApJ...752..137M} {752, 137}

\bibitem[\protect\citeauthoryear{{Offringa}, {Mertens}, {van der Tol},
  {Veenboer}, {Gehlot}, {Koopmans}  \& {Mevius}}{{Offringa}
  et~al.}{2019}]{2019A&A...631A..12O}
{Offringa} A.~R.,  {Mertens} F.,  {van der Tol} S.,  {Veenboer} B.,  {Gehlot}
  B.~K.,  {Koopmans} L.~V.~E.,   {Mevius} M.,  2019, \mn@doi [\aap]
  {10.1051/0004-6361/201935722}, \href
  {https://ui.adsabs.harvard.edu/abs/2019A&A...631A..12O} {631, A12}

\bibitem[\protect\citeauthoryear{Parsons \& Backer}{Parsons \&
  Backer}{2009}]{Parsons:2009ju}
Parsons A.~R.,  Backer D.~C.,  2009, \mn@doi [Astron. J.]
  {10.1088/0004-6256/138/1/219}, 138, 219

\bibitem[\protect\citeauthoryear{{Parsons}, {Pober}, {Aguirre}, {Carilli},
  {Jacobs}  \& {Moore}}{{Parsons} et~al.}{2012}]{2012ApJ...756..165P}
{Parsons} A.~R.,  {Pober} J.~C.,  {Aguirre} J.~E.,  {Carilli} C.~L.,  {Jacobs}
  D.~C.,   {Moore} D.~F.,  2012, \mn@doi [\apj] {10.1088/0004-637X/756/2/165},
  \href {https://ui.adsabs.harvard.edu/abs/2012ApJ...756..165P} {756, 165}

\bibitem[\protect\citeauthoryear{{Parsons} et~al.,}{{Parsons}
  et~al.}{2014}]{2014ApJ...788..106P}
{Parsons} A.~R.,  et~al., 2014, \mn@doi [\apj] {10.1088/0004-637X/788/2/106},
  \href {https://ui.adsabs.harvard.edu/abs/2014ApJ...788..106P} {788, 106}

\bibitem[\protect\citeauthoryear{{Parsons}, {Liu}, {Ali}  \& {Cheng}}{{Parsons}
  et~al.}{2016}]{2016ApJ...820...51P}
{Parsons} A.~R.,  {Liu} A.,  {Ali} Z.~S.,   {Cheng} C.,  2016, \mn@doi [\apj]
  {10.3847/0004-637X/820/1/51}, \href
  {https://ui.adsabs.harvard.edu/abs/2016ApJ...820...51P} {820, 51}

\bibitem[\protect\citeauthoryear{{Pascua} et~al.,}{{Pascua}
  et~al.}{2024}]{Pascua}
{Pascua} R.,  et~al., 2024, \mn@doi [arXiv e-prints]
  {10.48550/arXiv.2410.01872}, \href
  {https://ui.adsabs.harvard.edu/abs/2024arXiv241001872P} {p. arXiv:2410.01872}

\bibitem[\protect\citeauthoryear{{Patwa}, {Sethi}  \& {Dwarakanath}}{{Patwa}
  et~al.}{2021}]{2021MNRAS.504.2062P}
{Patwa} A.~K.,  {Sethi} S.,   {Dwarakanath} K.~S.,  2021, \mn@doi [\mnras]
  {10.1093/mnras/stab989}, \href
  {https://ui.adsabs.harvard.edu/abs/2021MNRAS.504.2062P} {504, 2062}

\bibitem[\protect\citeauthoryear{{Paul} et~al.,}{{Paul}
  et~al.}{2014}]{2014ApJ...793...28P}
{Paul} S.,  et~al., 2014, \mn@doi [\apj] {10.1088/0004-637X/793/1/28}, \href
  {https://ui.adsabs.harvard.edu/abs/2014ApJ...793...28P} {793, 28}

\bibitem[\protect\citeauthoryear{{Peckham}}{{Peckham}}{1973}]{1973MNRAS.165...25P}
{Peckham} R.~J.,  1973, \mn@doi [\mnras] {10.1093/mnras/165.1.25}, \href
  {https://ui.adsabs.harvard.edu/abs/1973MNRAS.165...25P} {165, 25}

\bibitem[\protect\citeauthoryear{{Pober} et~al.,}{{Pober}
  et~al.}{2014}]{2014ApJ...782...66P}
{Pober} J.~C.,  et~al., 2014, \mn@doi [\apj] {10.1088/0004-637X/782/2/66},
  \href {https://ui.adsabs.harvard.edu/abs/2014ApJ...782...66P} {782, 66}

\bibitem[\protect\citeauthoryear{{Pober} et~al.,}{{Pober}
  et~al.}{2016}]{2016ApJ...819....8P}
{Pober} J.~C.,  et~al., 2016, \mn@doi [\apj] {10.3847/0004-637X/819/1/8}, \href
  {https://ui.adsabs.harvard.edu/abs/2016ApJ...819....8P} {819, 8}

\bibitem[\protect\citeauthoryear{{Price}}{{Price}}{2016}]{2016ascl.soft03013P}
{Price} D.~C.,  2016, {PyGDSM: Python interface to Global Diffuse Sky Models},
  Astrophysics Source Code Library, record ascl:1603.013 (\mn@eprint {ascl}
  {1603.013})

\bibitem[\protect\citeauthoryear{{Pritchard} \& {Loeb}}{{Pritchard} \&
  {Loeb}}{2012}]{pritchard12}
{Pritchard} J.~R.,  {Loeb} A.,  2012, \mn@doi [Reports on Progress in Physics]
  {10.1088/0034-4885/75/8/086901}, \href
  {https://ui.adsabs.harvard.edu/abs/2012RPPh...75h6901P} {75, 086901}

\bibitem[\protect\citeauthoryear{{Robertson}}{{Robertson}}{2022}]{2022ARA&A..60..121R}
{Robertson} B.~E.,  2022, \mn@doi [\araa]
  {10.1146/annurev-astro-120221-044656}, \href
  {https://ui.adsabs.harvard.edu/abs/2022ARA&A..60..121R} {60, 121}

\bibitem[\protect\citeauthoryear{{Roshi} \& {Perley}}{{Roshi} \&
  {Perley}}{2003}]{2003ASPC..306..109R}
{Roshi} D.~A.,  {Perley} R.~A.,  2003, in {Minh} Y.~C.,  ed.,  Astronomical
  Society of the Pacific Conference Series Vol. 306, New technologies in VLBI.
  pp 109--121

\bibitem[\protect\citeauthoryear{{Rusholme}}{{Rusholme}}{2005}]{2005IAUS..201..512R}
{Rusholme} B.,  2005, in {Lasenby} A.~N.,  {Wilkinson} A.,  eds, ~ Vol. 201,
  New Cosmological Data and the Values of the Fundamental Parameters. p.~512

\bibitem[\protect\citeauthoryear{{Santos}, {Cooray}  \& {Knox}}{{Santos}
  et~al.}{2005}]{2005ApJ...625..575S}
{Santos} M.~G.,  {Cooray} A.,   {Knox} L.,  2005, \mn@doi [\apj]
  {10.1086/429857}, \href
  {https://ui.adsabs.harvard.edu/abs/2005ApJ...625..575S} {625, 575}

\bibitem[\protect\citeauthoryear{{Scott} \& {Rees}}{{Scott} \&
  {Rees}}{1990}]{1990MNRAS.247..510S}
{Scott} D.,  {Rees} M.~J.,  1990, \mnras, \href
  {https://ui.adsabs.harvard.edu/abs/1990MNRAS.247..510S} {247, 510}

\bibitem[\protect\citeauthoryear{{Shaw}, {Sigurdson}, {Pen}, {Stebbins}  \&
  {Sitwell}}{{Shaw} et~al.}{2014}]{2014ApJ...781...57S}
{Shaw} J.~R.,  {Sigurdson} K.,  {Pen} U.-L.,  {Stebbins} A.,   {Sitwell} M.,
  2014, \mn@doi [\apj] {10.1088/0004-637X/781/2/57}, \href
  {https://ui.adsabs.harvard.edu/abs/2014ApJ...781...57S} {781, 57}

\bibitem[\protect\citeauthoryear{{Shaw}, {Sigurdson}, {Sitwell}, {Stebbins}  \&
  {Pen}}{{Shaw} et~al.}{2015}]{2015PhRvD..91h3514S}
{Shaw} J.~R.,  {Sigurdson} K.,  {Sitwell} M.,  {Stebbins} A.,   {Pen} U.-L.,
  2015, \mn@doi [\prd] {10.1103/PhysRevD.91.083514}, \href
  {https://ui.adsabs.harvard.edu/abs/2015PhRvD..91h3514S} {91, 083514}

\bibitem[\protect\citeauthoryear{{Subramanian} \& {Padmanabhan}}{{Subramanian}
  \& {Padmanabhan}}{1993}]{1993MNRAS.265..101S}
{Subramanian} K.,  {Padmanabhan} T.,  1993, \mn@doi [\mnras]
  {10.1093/mnras/265.1.101}, \href
  {https://ui.adsabs.harvard.edu/abs/1993MNRAS.265..101S} {265, 101}

\bibitem[\protect\citeauthoryear{{Tan} et~al.,}{{Tan}
  et~al.}{2021}]{2021ApJS..255...26T}
{Tan} J.,  et~al., 2021, \mn@doi [\apjs] {10.3847/1538-4365/ac0533}, \href
  {https://ui.adsabs.harvard.edu/abs/2021ApJS..255...26T} {255, 26}

\bibitem[\protect\citeauthoryear{{Thyagarajan} et~al.,}{{Thyagarajan}
  et~al.}{2015}]{2015ApJ...807L..28T}
{Thyagarajan} N.,  et~al., 2015, \mn@doi [\apjl] {10.1088/2041-8205/807/2/L28},
  \href {https://ui.adsabs.harvard.edu/abs/2015ApJ...807L..28T} {807, L28}

\bibitem[\protect\citeauthoryear{{Watson} et~al.,}{{Watson}
  et~al.}{2003}]{2003MNRAS.341.1057W}
{Watson} R.~A.,  et~al., 2003, \mn@doi [\mnras]
  {10.1046/j.1365-8711.2003.06338.x}, \href
  {https://ui.adsabs.harvard.edu/abs/2003MNRAS.341.1057W} {341, 1057}

\bibitem[\protect\citeauthoryear{{Wilensky}, {Brown}  \& {Hazelton}}{{Wilensky}
  et~al.}{2023}]{2023MNRAS.521.5191W}
{Wilensky} M.~J.,  {Brown} J.,   {Hazelton} B.~J.,  2023, \mn@doi [\mnras]
  {10.1093/mnras/stad863}, \href
  {https://ui.adsabs.harvard.edu/abs/2023MNRAS.521.5191W} {521, 5191}

\bibitem[\protect\citeauthoryear{{Wolz}, {Abdalla}, {Blake}, {Shaw}, {Chapman}
  \& {Rawlings}}{{Wolz} et~al.}{2014}]{2014MNRAS.441.3271W}
{Wolz} L.,  {Abdalla} F.~B.,  {Blake} C.,  {Shaw} J.~R.,  {Chapman} E.,
  {Rawlings} S.,  2014, \mn@doi [\mnras] {10.1093/mnras/stu792}, \href
  {https://ui.adsabs.harvard.edu/abs/2014MNRAS.441.3271W} {441, 3271}

\makeatother
\end{thebibliography}



\appendix


\section{Examples of DPSS used in the mainlobe filter}
\label{app1}

\begin{figure*}
    \centering
         \includegraphics[width=\textwidth]{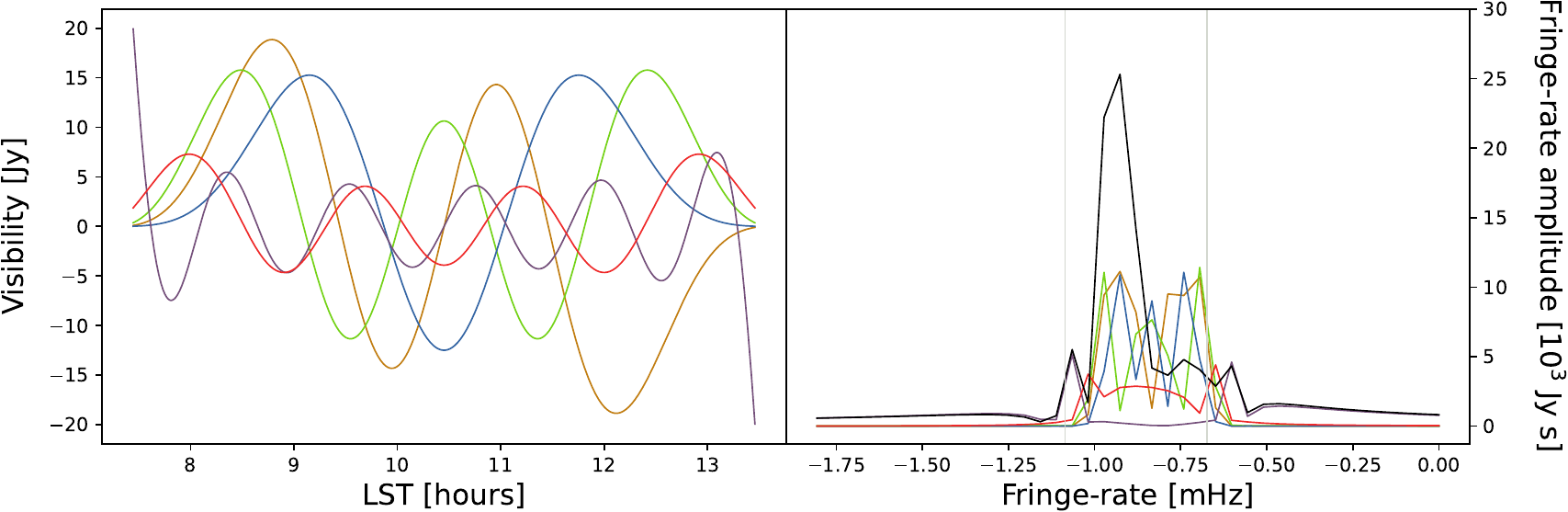}
     \caption{Examples of DPSS used for mainlobe filtering. Five DPSS are shown out of a set of 18 that are used for the complete mainlobe filter. The DPSS are shown on the left, and their corresponding fringe-rates on the right. The black line shows the amplitude of the 5 summed DPSS in fringe-rate space.}
     \label{dpss}
\end{figure*}

\begin{figure*}
    \centering
         \includegraphics[width=\textwidth]{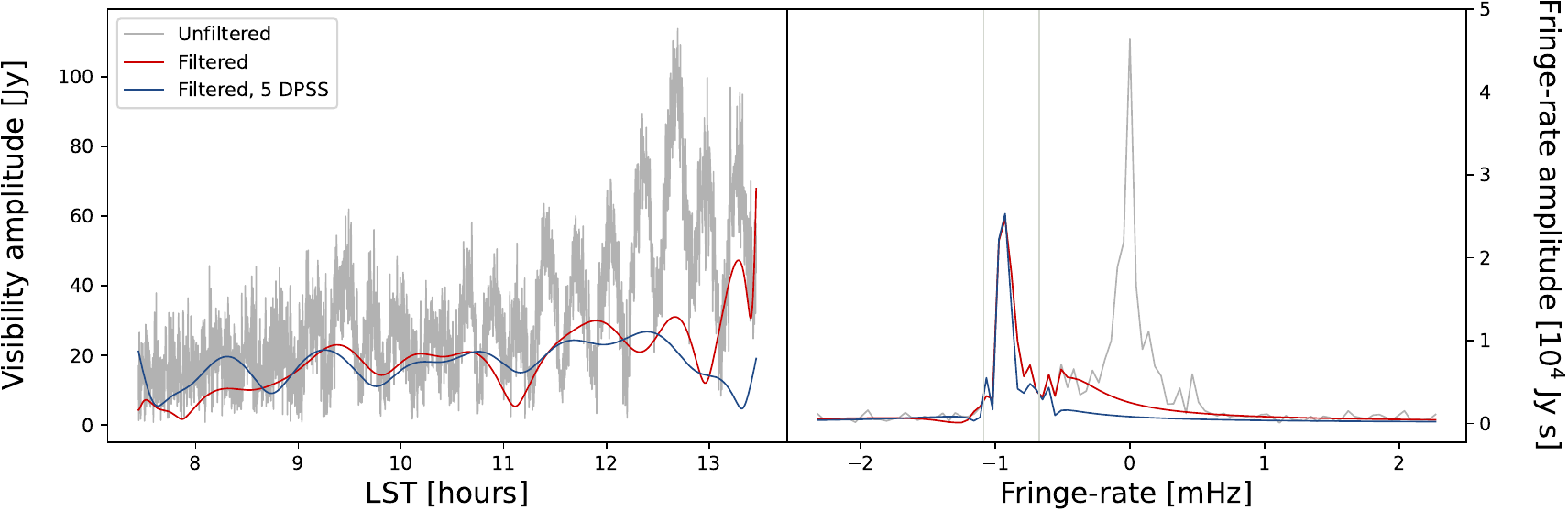}
     \caption{The example visibility data and its fringe-rates, before and after mainlobe filtering. In the left panel, the input visibilities are shown in grey, the mainlobe-filtered visibilities are shown in red, and the filtered visibilities using only 5 DPSS  are shown in blue, as described in the text. The right panel shows the corresponding fringe-rates.}
     \label{filt5}
\end{figure*}

\begin{figure*}
\includegraphics[width=\textwidth]{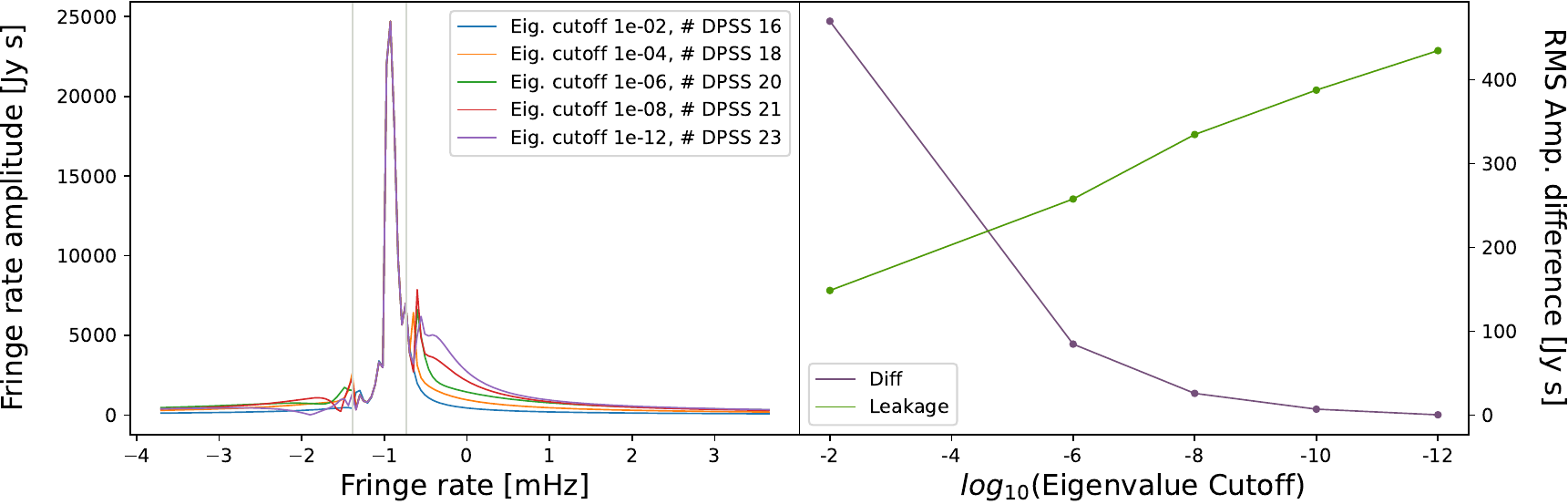}
     \caption{The results of an experiment to determine the consequences of different eigenvalue cutoffs when selecting DPSS for a bandpass (mainlobe) filter. The left panel shows the fringe-rates after filtering the visibilities for baseline (3, 6) at 100~MHz with different cutoffs as in the legend. The right panel shows the difference between the filtered and input fringe-rates, and the leakage outside the filter bounds, by cutoff. How these are calculated  is defined in the text.}
     \label{eigenval}
\end{figure*}

We will demonstrate filtering of  the  visibilities that are shown in the top row of Fig. \ref{mainlobe_vis_summary} except that they have not been coherently averaged.

 Given a tophat filter with a width and a center in fringe-rate space, a set of DPSS are generated that have power concentrated within the tophat filter bounds. The mechanism for generating a set involves  specifying parameters such as length, resolution, and filter requirements, then selecting DPSS that most concentrate power within the filter bounds. The selection method is based on eigenvalue cutoff, which is described in Appendix \ref{app2}. Eighteen DPSS are selected to use for fitting the example visibilities; the 5 DPSS with greatest amplitude after fitting are shown in the left panel of Fig. \ref{dpss}. These are real valued, with power in fringe-rate space that is centered at fringe-rate 0 mHz, but before fitting are shifted to the centre  of the tophat filter by converting to complex values and assigning appropriate phases. The right panel of Fig. \ref{dpss} shows the amplitudes of the fringe-rates for the 5 DPSS after rephasing (colored, corresponding to those in the left panel), and the black line shows the amplitude of their (complex) sum. The vertical grey lines indicate the bounds of the tophat filter.

 Fig. \ref{filt5} places these 5 DPSS within the context of the mainlobe  filtering of the chosen visibilities over time, shown in grey in the left panel. The filtered visibilities using all 18 DPSS are shown in red and the filtered visibilities using only the 5 DPSS are shown in blue. The right panel shows the fringe-rates of the visibilities in the left panel.

\section{DPSS eigenvalue cutoff}
\label{app2}    

DPSS arise from a system of equations whose solutions have a Fourier spectrum concentrated within certain bounds \citep{330397}. The system of equations can be expressed as a matrix equation where the DPSS are eigenvectors. Since they are eigenvectors, they have an associated eigenvalue, with larger eigenvalues indicating a more concentrated spectrum within the filter bounds. By selecting DPSS with eigenvalues above a certain value, the ``eigenvalue cutoff'', a limited set of DPSS can be obtained and used to bandpass filter a signal.

One then has to select an eigenval cutoff, and there is a trade-off between how well the spectrum is concentrated, and how much spectral leakage there is outside the bounds. Fig. \ref{eigenval} demonstrates this trade-off. We take a time sequence of simulated visibilities for baseline (3, 6) at 100~MHz, and filter it using DPSS, with 5 different eigenvalue cutoffs from $10^{-2}$ to $10^{-12}$. The left panel shows the fringe-rates after filtering, with the legend indicating the eigenvalue cutoff and the number of DPSS that were selected by that cutoff. The right panel shows how well the filter has performed, using two measures: 1. The RMS difference between the fringe-rates of the input visibilities and the fringe-rates of the filtered visibilities, within the filter bounds, and 2. the RMS difference between the fringe-rates of the filtered visibilities and 0, outside the bounds. In other words, 1. indicates how well the filtered spectrum matches the input spectrum within the bounds, and 2. indicates how much spectral leakage there is. In the legend, ``Diff'' denotes measure 1, and ``Leakage'' denotes measure 2. We want both measures to be small. The filter bounds are indicated by the vertical grey lines. 

As the eigenvalue cutoff value is reduced, more DPSS modes are selected, and the fit of the filtered spectrum within the filter bound improves (``Diff'' goes to 0), but the leakage gets worse (``Leakage'' increases).  However, it is clear that some leakage will always be present, whereas some ``Diff'' will not always be present, since it is possible to reduce ``Diff'' to near 0. In the case of the mainlobe filter, this means that the fringe rates we want to retain are preserved very well, as is seen in the second row, left panel,  of Fig. \ref{mainlobe_vis_summary}. The notch filter is a different matter, where the match is not as good, but the notch filter suffers from issues of resolution of fringe-rates where the notch is so small, and this needs to be further investigated.

In this work we choose to minimize ``Diff'', and use an eigenvalue cutoff of $10^{-12}$, accepting that there will always be leakage outside the filter bounds. The actual form of this leakage can be examined in further work, in particular how the RMS difference is spread across the fringe-rates outside the bounds, and the impact of that. Within the filter bounds, at least for the mainlobe filter, we retain a good match to the signal. In the next section we will show the effect of different cutoffs on power spectra, and that the effect is limited.

Note that we have examined the consequences of different cutoffs for only one baseline. The same analysis could be made for all baselines, with cutoffs and the number of DPSS modes varying by baseline. The effect of the cutoff, and the filters in general, should also be examined for real data, but the cutoff of $10^{-12}$ is a staring point that allows inspection of the filter effect for the mainlobe filter within the filter bounds, and from that the cutoff can be adjusted to observe its impact.

\subsection{The cutoff's effect on wedge power spectra}

\begin{figure*}
    \centering
         \includegraphics[width=\textwidth]{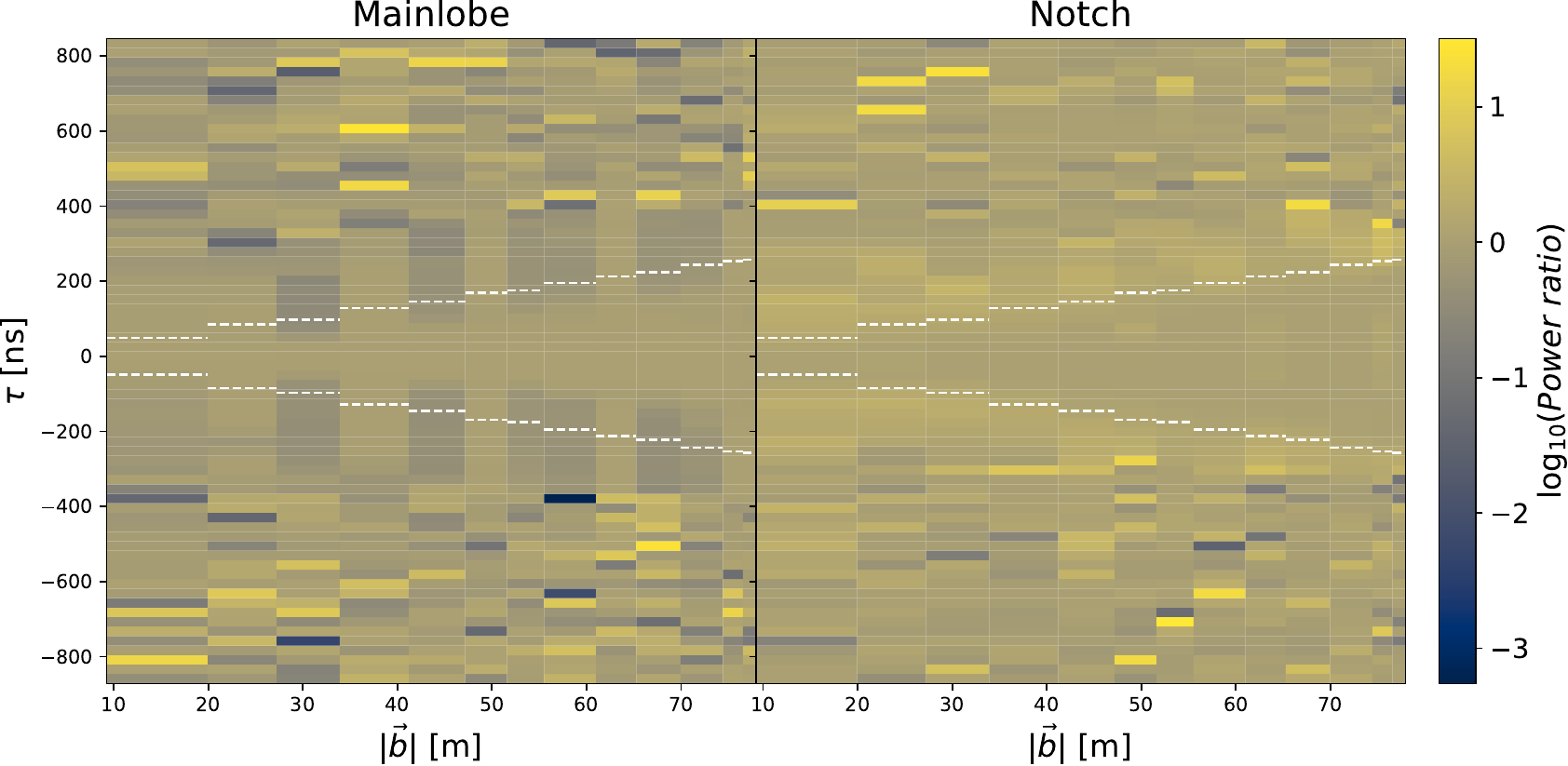}
     \caption{Each of these power spectra is the result of dividing power spectra generated from visibilities filtered using different eigenvalue cutoffs. See text.}
     \label{spectra_different_cutoff}
\end{figure*}

\begin{figure*}
    \centering
         \includegraphics[width=\textwidth]{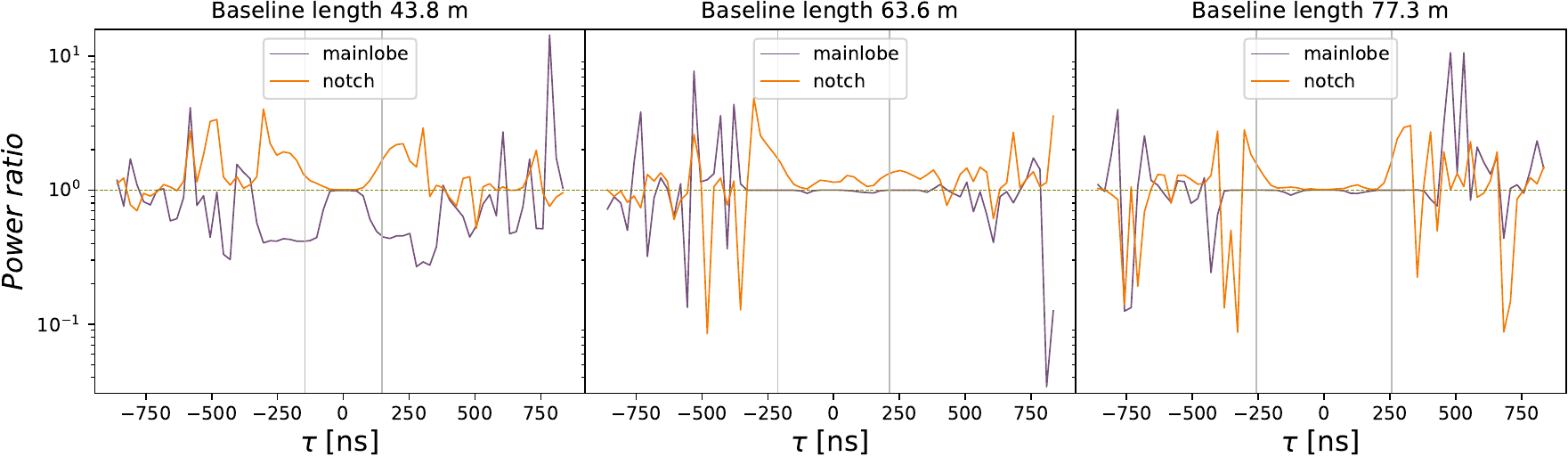}
     \caption{Vertical cuts through the power spectra in Fig. \ref{spectra_different_cutoff}.}
     \label{av_eig_wedges_ratio_cuts}
\end{figure*}

To compare power spectra filtered with different eigenvalue cutoffs, we use the ratio plots in Fig. \ref{spectra_different_cutoff}.
These ratio plots have a different meaning to the ratio plots elsewhere in this paper.  Fig. \ref{wedges} contains power spectra from visibilities passed through the mainlobe and notch filters using an eigenvalue cutoff of $10^{-12}$. The same power power spectra were regenerated using an eigenvalue cutoff of $10^{-6}$. Then, the latter were divided by the former.

The plots show that a cutoff of $10^{-6}$ can increase the power by a factor of $\sim 10$ in some locations of the power spectra. It can also decrease the power by a factor of $\sim 10^{-3}$ in other locations, although this happens less often. Fig. \ref{av_eig_wedges_ratio_cuts} shows that most of the locations where there are substantial increases/decreases lie outside the wedge; inside the wedge there is little change in power. The changes are also positive and negative, rather than an obvious increase or decrease over the entire spectrum.

While this does not indicate which cutoff value is ``best'', for example for detecting the EoR, it does show that a $10^{-6}$ change in magnitude of the cutoff does not radically alter a power spectrum. Given that there are so many other data manipulations involved -- coherent and incoherent time averaging, redundant baseline averaging  -- it is possible that the cutoff value is of minor concern in comparison.

\section{Author Affiliations}\label{sec:affiliations}
\noindent\textit{
$^{1}$ Queen Mary University London, London E1 4NS, UK\\
$^{2}$ Jodrell Bank Centre for Astrophysics, University of Manchester, Manchester, M13 9PL, United Kingdom\\
$^{3}$ Department of Physics and Astronomy,  University of Western Cape, Cape Town, 7535, South Africa\\
$^{4}$ Department of Astronomy, University of California, Berkeley, CA\\
$^{5}$ South African Radio Astronomy Observatory, Black River Park, 2 Fir Street, Observatory, Cape Town, 7925, South Africa\\
$^{6}$ Department of Physics and Astronomy, University of Pennsylvania, Philadelphia, PA\\
$^{7}$ Cavendish Astrophysics, University of Cambridge, Cambridge, UK\\
$^{8}$ School of Earth and Space Exploration, Arizona State University, Tempe, AZ\\
$^{9}$ Department of Physics, Winona State University, Winona, MN\\
$^{10}$ INAF-Istituto di Radioastronomia, via Gobetti 101, 40129 Bologna, Italy\\
$^{11}$ Department of Physics and Electronics, Rhodes University, PO Box 94, Grahamstown, 6140, South Africa\\
$^{12}$ National Radio Astronomy Observatory, Charlottesville, VA\\
$^{13}$ National Radio Astronomy Observatory, Socorro, NM 87801, USA\\
$^{14}$ MIT Kavli Institute, Massachusetts Institute of Technology, Cambridge, MA\\
$^{15}$ Department of Physics, Massachusetts Institute of Technology, Cambridge, MA\\
$^{16}$ Centre for Strings, Gravitation and Cosmology, Department of Physics, Indian Institute of Technology Madras, Chennai 600036, India\\
$^{17}$ Radio Astronomy Lab, University of California, Berkeley, CA\\
$^{18}$ Department of Physics, University of California, Berkeley, CA\\
$^{19}$ Department of Physics and Astronomy, University of California, Los Angeles, CA\\
$^{20}$ National Radio Astronomy Observatory, Socorro, NM\\
$^{21}$ Department of Physics and McGill Space Institute, McGill University, 3600 University Street, Montreal, QC H3A 2T8, Canada\\
$^{22}$ School of Physics, University of Melbourne, Parkville, VIC 3010, Australia\\
$^{23}$ Department of Physics, University of Washington, Seattle, WA\\
$^{24}$ eScience Institute, University of Washington, Seattle, WA\\
$^{\dagger}$ NASA Hubble Fellow\\
$^{25}$ Department of Physics, Brown University, Providence, RI\\
$^{26}$ Department of Physics, Stellenbosch University, Matieland 7602, South Africa\\
$^{27}$ Scuola Normale Superiore, 56126 Pisa, PI, Italy\\
$^{28}$ International Centre for Radio Astronomy Research, Curtin University, Bentley, WA 6102, Australia\\
$^{29}$ ARC Centre of Excellence for All Sky Astrophysics in 3 Dimensions (ASTRO 3D), Bentley, WA 6102, Australia\\
$^{30}$ School of Physics, University of Melbourne, Parkville, VIC 3010 Australia\\
$^{31}$ Raman Research Institute\\
$^{32}$ Commonwealth Scientific and Industrial Research Organisation (CSIRO), Space \& Astronomy, P. O. Box 1130, Bentley, WA 6102, Australia\\
$^{33}$ Center for Astrophysics, Harvard \& Smithsonian, Cambridge, MA\\
$^{34}$ American Astronomical Society, Washington, DC
}

\label{lastpage}
\bsp	
\end{document}